\documentclass[final, 1p, times]{elsarticle}

\usepackage{amssymb}
\usepackage{amsmath}
\usepackage[colorlinks=true, linkcolor=blue, citecolor=blue, urlcolor=blue]{hyperref}
\usepackage{longtable}
\usepackage{array}
\usepackage{booktabs}
\usepackage{xcolor}
\usepackage{geometry}
\usepackage{pdflscape}
\usepackage{graphicx}
\usepackage{footnote}
\usepackage{fontenc}
\usepackage{caption}
\usepackage{float}
\usepackage{placeins}
\usepackage{tabularx}
\usepackage{multirow}
\usepackage{comment}

\journal{}

\graphicspath{{figures/}}

\begin{document}

\begin{frontmatter}

\title{\texorpdfstring{Lessons learned from field demonstrations of model predictive control and reinforcement learning for residential and commercial HVAC: A review}{Lessons learned from field demonstrations of model predictive control and reinforcement learning for residential and commercial HVAC: A review}}

\author[1]{Arash J. Khabbazi}
\ead{arashjkh@purdue.edu}

\author[1,5]{Elias N. Pergantis}
\ead{elias.pergantis@tranetechnologies.com}

\author[1]{Levi D. Reyes Premer}
\ead{lreyespr@purdue.edu}

\author[1]{Panagiotis Papageorgiou}
\ead{ppapageo@purdue.edu}

\author[1]{Alex H. Lee}
\ead{lee3742@purdue.edu}

\author[1]{James E. Braun}
\ead{jbraun@purdue.edu}

\author[2,3]{Gregor P. Henze}
\ead{gregor.henze@colorado.edu}

\author[1,4]{Kevin J. Kircher}
\ead{kkirche@purdue.edu}

\affiliation[1]{organization={School of Mechanical Engineering, Purdue University},
city={West Lafayette, IN},
country={USA}}

\affiliation[2]{organization={Department of Civil, Environmental and Architectural Engineering, University of Colorado},
city={Boulder, CO},
country={USA}}

\affiliation[3]{organization={National Renewable Energy Laboratory},
city={Golden, CO},
country={USA}}

\affiliation[4]{organization={Elmore Family School of Electrical and Computer Engineering, Purdue University},
city={West Lafayette, IN},
country={USA}}

\affiliation[5]{organization={Trane Technologies, Residential R\&D Group},
city={Tyler, TX},
country={USA}}

\begin{abstract}
A large body of simulation research suggests that model predictive control (MPC) and reinforcement learning (RL) for heating, ventilation, and air-conditioning (HVAC) in residential and commercial buildings could reduce energy costs, pollutant emissions, and strain on power grids. Despite this potential, neither MPC nor RL has seen widespread industry adoption. Field demonstrations could accelerate MPC and RL adoption by providing real-world data that support the business case for deployment. Here we review 24 papers that document field demonstrations of MPC and RL in residential buildings and 80 in commercial buildings. After presenting demographic information -- such as experiment scopes, locations, and durations -- this paper analyzes experiment protocols and their influence on performance estimates. We find that 71\% of the reviewed field demonstrations use experiment protocols that may lead to unreliable performance estimates. Over the remaining 29\% that we view as reliable, the weighted-average cost savings, weighted by experiment duration, are 16\% in residential buildings and 13\% in commercial buildings. While these savings are potentially attractive, making the business case for MPC and RL also requires characterizing the costs of deployment, operation, and maintenance. Only 13 of the 104 reviewed papers report these costs or discuss related challenges. Based on these observations, we recommend directions for future field research, including: Improving experiment protocols; reporting deployment, operation, and maintenance costs; designing algorithms and instrumentation to reduce these costs; controlling HVAC equipment alongside other distributed energy resources; and pursuing emerging objectives such as peak shaving, arbitraging wholesale energy prices, and providing power grid reliability services.
\end{abstract}

\begin{graphicalabstract}
\includegraphics[width=0.975\textwidth]{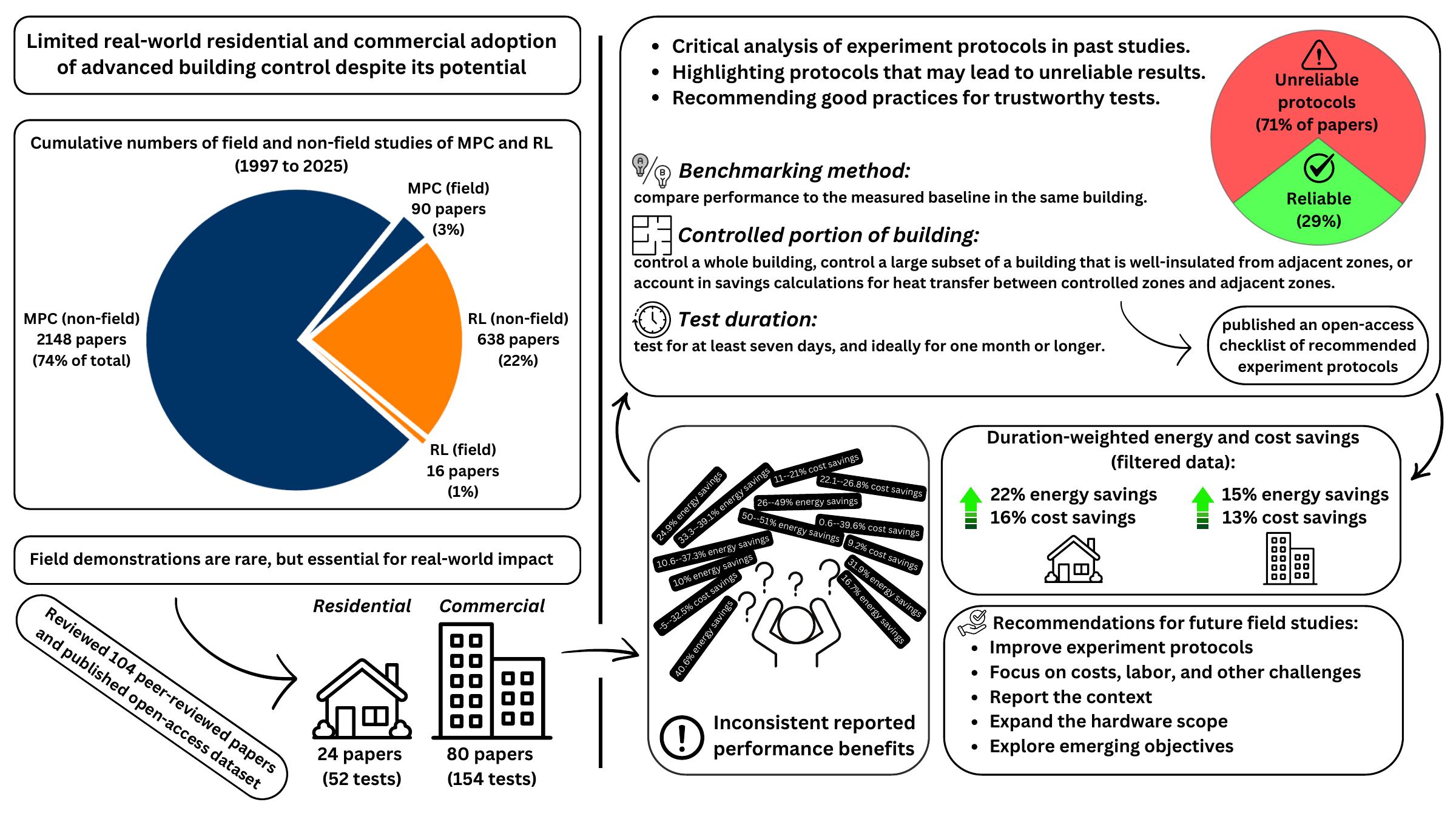}
\end{graphicalabstract}

\begin{highlights}
\item Review of MPC and RL field work on heating, ventilation, and air conditioning
\item Lessons learned from 206 tests in residential and commercial buildings
\item Critical analysis of experiment protocols in past field work
\item Recommendations for future field work and algorithm development
\item Open dataset with paper details and experiment checklist
\end{highlights}


\end{frontmatter}

%% main text %%%%%%%%%%%%%%%%%%%%%%%%%%%%%%%%%%%%%%%%%%%%%%%%%%%%%%%%%%%%%%%%%%%%%%%%%%%%%%%%%%%%%%%%%%%%%%%%%%

\newpage
\section{Introduction}

Heating, ventilation, and air conditioning (HVAC) systems in residential and commercial buildings cause about 15\% of global greenhouse gas emissions\footnote{From \cite{IEA2022_heating}, heating residential and commercial buildings accounts for 10\% of global greenhouse gas emissions. From \cite{IEA2022_cooling} and \cite{Trane2023}, cooling contributes 5\% (3.2\% from energy use and 1.8\% from refrigerants).} and cost on the order of \$1 trillion per year in energy bills\footnote{From \cite{urge2015heating}, residential and commercial buildings use about 32\% of global annual final energy, of which HVAC systems use about 57\% (or about 18\% of the global total). From \cite{enerdata}, global annual final energy expenditures are about \$10 trillion (inflation-adjusted to 2024 dollars). A rough estimate of annual global expenditures on HVAC energy is therefore \$1.8 trillion.}. Over the last two decades, researchers have shown that advanced control methods -- such as model predictive control (MPC) and reinforcement learning (RL) -- can significantly mitigate greenhouse gas emissions, reduce energy costs, and turn HVAC systems into active participants in the power grid. Despite these research efforts, however, HVAC industry adoption of MPC and RL remains slow due to market barriers, deployment challenges, and other factors \cite{Khabbazi2024-im, Henze2024-qa}. This paper aims to move the technology closer to real-world adoption at scale by (1) critically reviewing what the research community has learned from field demonstrations of MPC and RL and (2) highlighting important directions for future work.

Advanced HVAC control research has at least 35 years of history. In the 1990s and early 2000s, studies mainly focused on open-loop optimal control systems, which plan future control actions and implement them without feedback from real-time measurements or updated forecasts \cite{Braun1990-sn, Henze2003-qh}. Open-loop optimal control concepts were later extended to the framework of MPC \cite{Henze1997-ss}. MPC plans control actions over a receding forecast horizon, executes the first planned action, updates measurements and forecasts, then repeats the process at the next time step. While MPC uses predictive models to optimize control actions, RL is a potentially model-free alternative that was first implemented for HVAC systems in  \cite{Mozer-1998-1st-ever-RL, Henze2003-ig}. Instead of directly using a mathematical model, RL can learn a control policy from real-world data or data from model-based simulations. This means that RL may be useful in cases in which developing models would be costly or impractical. In a parallel line of work, data-enabled predictive control (DeePC) has emerged as a model-free variant of MPC \cite{coulson2019data}. DeePC applies Willems' Fundamental Lemma \cite{Willems2013-jj} of behavioral systems theory to predict system responses directly from data, eliminating the challenge of developing an explicit system model. This makes DeePC potentially attractive for HVAC systems with complex and dynamic characteristics that make mathematical modeling difficult.

\begin{figure*}[!htbp]
    \centering
    \includegraphics[width=0.99\textwidth]{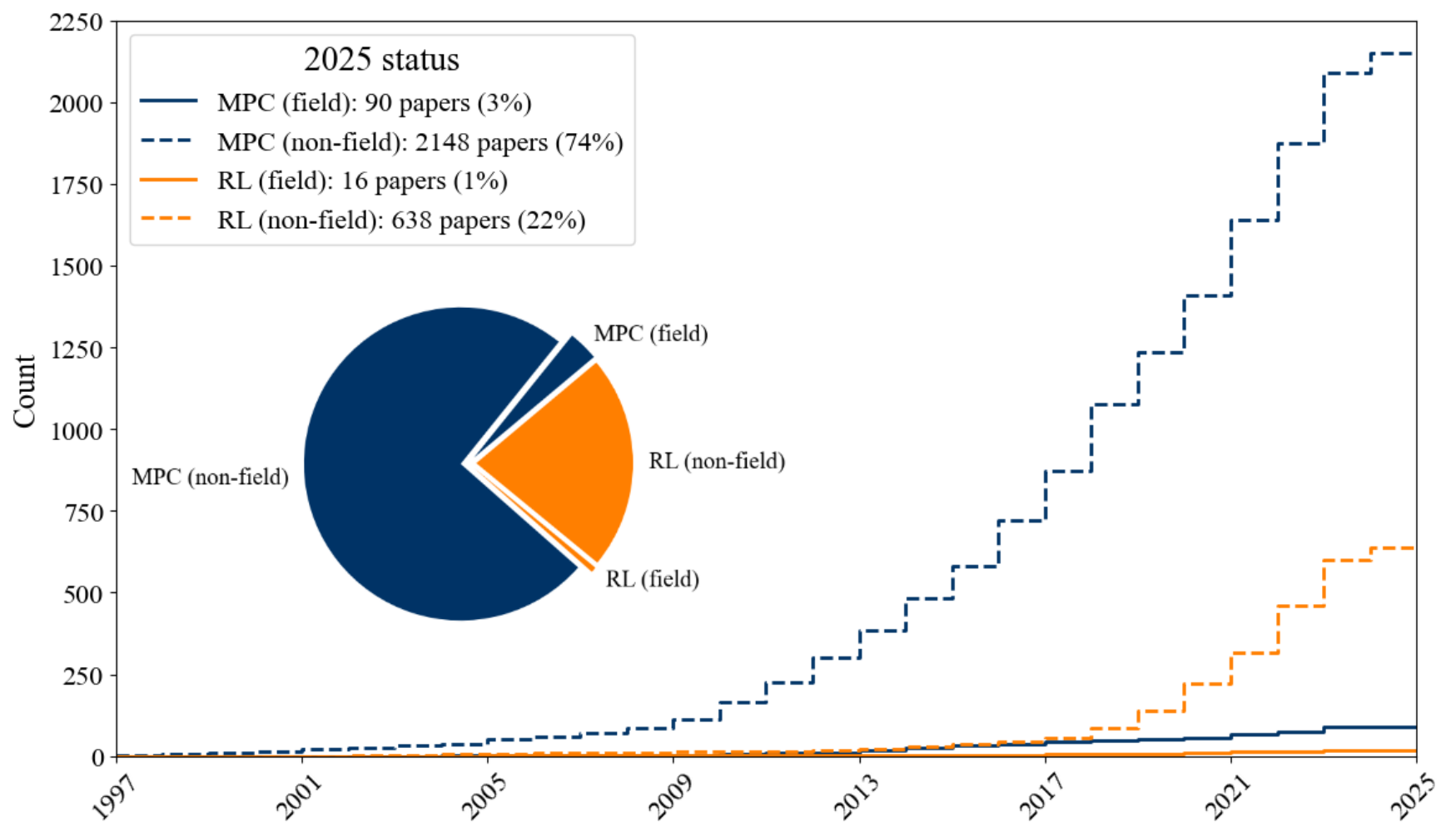}
    \caption{Cumulative numbers of field and non-field studies of MPC and RL. MPC research accelerated around 2010, while RL research accelerated about a decade later. Across both algorithms, field studies comprise just 4\% of all published papers. Two of the 104 papers reviewed here demonstrated both MPC and RL.}
    \label{FIG1}
\end{figure*}

Research on MPC and RL for HVAC control has grown significantly in recent years, but experimental research remains limited. \autoref{FIG1} shows the cumulative growth in field and non-field papers on MPC and RL since 1997\footnote{To create this figure, we used a comprehensive Scopus search with a wide array of keywords related to MPC, RL, HVAC, and buildings. The period covered for MPC begins in 1997 with the first journal paper on MPC \cite{Henze1997-ss}. For RL, it begins in 2003 with the first RL paper \cite{Henze2003-ig}. The ending year is the time of submission of this paper.}. Only 4\% of 2,892 relevant papers focus on field demonstrations. This limited focus is most likely due to the costs, time constraints, and other challenges associated with field demonstrations, such as gaining data access and control authority in real buildings. Despite these obstacles, field demonstrations could show the practical benefits of MPC and RL to key stakeholders, such as decision-makers in HVAC companies and government funding agencies \cite{De_Chalendar2024-la}. Field demonstrations also help identify unforeseen implementation challenges and improve our understanding of how advanced HVAC control systems handle the real-world uncertainties that simulation studies often overlook \cite{Henze2024-qa}.

Despite the importance of field demonstrations in advancing real-world adoption, the research community currently lacks a systematic and comprehensive review of field demonstrations. Previous efforts have provided limited coverage. For example, Sturzenegger et al. \cite{Sturzenegger2016-vd} and Blum et al. \cite{Blum2022-dc} reviewed small numbers of papers on field demonstrations of MPC in commercial buildings, covering 10 and 17 papers, respectively. Sturzenegger et al. and Blum et al. provided these partial reviews as context for their own field developments of MPC systems. This paper's authors recently reviewed field demonstrations of MPC and RL in commercial buildings, covering 56 field tests from 36 papers from 2005 to 2024, and reported ongoing deployment challenges \cite{Khabbazi2024-im}. Pergantis et al. \cite{Pergantis2024a} surveyed 12 papers on field demonstrations of MPC and RL in residential buildings in a paper that also developed and field-tested their own MPC system. More recently, Saloux \cite{Saloux2025-yt} carried out a broader partial review of MPC field demonstrations, focusing on methods to estimate energy savings from field implementations and referencing about 40 papers on residential and commercial buildings. Section \ref{sec:relatedWork} delves deeper into these five studies, as well as other relevant review papers on general HVAC control strategies, MPC, and RL. Building on past efforts, this paper provides the most comprehensive review to date, evaluating all peer-reviewed field demonstrations of which the authors are aware. In addition to reviewing previous publications, this paper looks at emerging trends and patterns in field demonstrations and proposes experiment protocols and guidelines to support reliable and scalable implementation of advanced HVAC control.

\subsection{Contributions of this paper}

This paper makes four main contributions to the literature on MPC and RL for residential and commercial HVAC. 

\begin{enumerate}

    \item \textbf{Comprehensive review:} This paper is the first systematic review of all known peer-reviewed field demonstrations of MPC and RL for residential and commercial HVAC.

    \item \textbf{Open-access dataset:} To aid future research, we provide an open-access dataset in \cite{field-demonstrations-HVAC-MPC-RL} containing the details of all field demonstrations reviewed in this paper.

    \item \textbf{Experiment protocols:} This paper critically analyzes experiment protocols used in past field studies, highlights protocols that may lead to unreliable results, and recommends good practices for trustworthy field studies.

    \item \textbf{Recommendations for future work:} Based on gaps identified in our review of 104 field studies, as well as the authors' collective experience conducting 21 field studies, we recommend directions for future HVAC control research. We also provide a checklist of recommended experiment protocols for future field demonstrations \cite{field-demonstrations-HVAC-MPC-RL}.
    
\end{enumerate}

\subsection{Organization of this paper}

This paper is structured as follows. Section \ref{sec:relatedWork} compares this paper to previous review papers on general HVAC control strategies, MPC, RL, and field demonstrations. Section \ref{sec:methodology} discusses the methodology, which includes the search strategy, savings evaluation approach, and statistical methods used. Section \ref{sec:demographics} presents demographic information, including locations, weather conditions, and experiment scopes. Section \ref{sec:protocols} highlights several important experiment protocol choices -- including benchmarking methods, controlled building portions, and test durations -- and critically evaluates the reliability of existing experimental work. Section \ref{sec:discussion} discusses electricity rate structures, building properties, control architectures, weather, and implementation challenges. This section also provides recommendations for future field studies and highlights limitations that future literature reviews could address. Section \ref{sec:conclusion} concludes the paper. \ref{nomenclature} summarizes the nomenclature used in this paper. \ref{sec:appendixa} contains two summary tables that present an overview of all the residential (\autoref{tab:res}) and commercial (\autoref{tab:com}) field demonstrations reviewed here.

\section{Related literature reviews}
\label{sec:relatedWork}

In recent years, a growing number of literature review papers have analyzed the application of advanced control -- including MPC, RL, and other methods -- to HVAC systems in residential and commercial buildings. This section summarizes these literature review papers and categorizes them according to their focus. To establish context, we begin with a brief discussion of literature reviews on general HVAC control strategies before delving into those that specifically address MPC, RL, and field demonstrations.

\subsection{General HVAC control reviews} %%%%%%% cited 18 papers

Early reviews on HVAC control focused primarily on optimal and supervisory control. Wang and Ma \cite{Wang2008-dv} reviewed optimal and supervisory HVAC control, early model-based approaches, and broader classifications of advanced building controls. With a particular focus on multi-agent systems, Dounis and Caraiscos \cite{Dounis2009-vw} surveyed advanced building controls for energy efficiency and occupant comfort. In \cite{Naidu2011-hard}, Naidu and Rieger reviewed `hard' control approaches, including proportional-integral-derivative (PID), optimal, model predictive, robust, nonlinear, and adaptive, while in \cite{Naidu2011-soft}, they reviewed `soft' control techniques, including artificial neural networks (ANNs), fuzzy logic, genetic algorithms, as well as hybrids of `hard' and `soft' control. Mirinejad et al. \cite{Mirinejad2012-ko} reviewed intelligent control strategies, with a particular focus on neuro-fuzzy and genetic-fuzzy approaches. In a comprehensive review of optimal control systems for energy and comfort management in smart buildings, Shaikh et al. \cite{Shaikh2014-vl} highlighted a variety of control methods, energy and comfort related trends, and future directions. Royapoor et al. \cite{Royapoor2018-ln} reviewed building control methods, ranging from conventional proportional-integral-derivative control to emerging methods like fuzzy logic and ANNs. Their review addressed challenges in adopting RL and integrating renewable electricity generation into power systems.

Occupant-centric control (OCC) has drawn recent attention in the context of building control. OCC integrates real-time occupancy data and occupant comfort feedback into control algorithms. Naylor et al. \cite{Naylor2018-td} provided a structured survey of OCC strategies and focused on the relevant sensing technologies and categorizing implementation approaches into different groups. Jung and Jazizadeh \cite{Jung2019-lf} conducted a performance-based analysis of OCC systems and provided a taxonomy to evaluate occupancy- and comfort-driven approaches. Focusing on field implementations, Park et al. \cite{Park2019-ul} reviewed practical challenges in deploying OCC strategies and discussed future directions. According to a review by Liu et al. \cite{Liu2023-by}, OCC is a transformative strategy which includes occupant presence, preferences, and interactions to enhance HVAC operation for energy efficiency and thermal comfort.

Other review papers have focused on optimization-based strategies in order to improve system performance and decision-making. An early review by Wang et al. \cite{Wang2009-rq} focused on multi-criteria decision-making approaches to assess renewables across different sustainability dimensions. Banos et al. \cite{Banos2011-ie} reviewed computational optimization strategies applied to different energy sectors. According to a review by Evins \cite{Evins2013-wp}, computational optimization, particularly multi-objective methods, improves sustainable building design in envelopes, building control, and renewables. Focusing on commercial buildings, Lazos et al. \cite{Lazos2014-go} reviewed the role of weather forecasts in optimal operation of energy management systems, focusing on load and generation predictions, and outlined an integrated framework for better control and energy savings. Focusing on the residential demand-side management, Esther and Kumar \cite{Esther2016-yl} presented a comprehensive review of various optimization approaches, such as deterministic, stochastic, and game-theoretic models. In addition, Aste et al. \cite{Aste2017-al} presented a framework for optimal building automation, which aims to improve decision-making under uncertainty and close the performance gap. More recently, 82 multi-objective studies were covered by Al Mindeel et al. \cite{Al-Mindeel2024-os}, targeting energy savings, thermal comfort, and indoor air quality.

\subsection{Reviews of MPC for buildings} %%%%%%%%% 18 papers cited - verified

Since the early 21st century, research on MPC for buildings has grown, with authors performing reviews with different scopes, some focusing only on MPC while others addressed it alongside other advanced control techniques. Henze \cite{Henze2013-gk} provided an early perspective on MPC capabilities in buildings and revealed its potential to significantly impact energy management and optimize intelligent building operations. One of the most comprehensive early reviews on MPC is that of Afram and Janabi-Sharifi \cite{Afram2014-rx}, which presented a systematic framework for MPC implementation in HVAC systems, from classifications and performance factors to practical applications. Killian and Kozek \cite{Killian2016-pm} posed ten questions about MPC implementation in building control, studying its challenges, potential, and future prospects, with a focus on practical steps for real-world application. Serale et al. \cite{Serale2018-ph} developed a detailed framework for MPC in buildings and presented a common taxonomy by addressing challenges in defining control problems, while demonstrating its potential to improve energy efficiency. Drgoňa et al. \cite{Drgona2020-fh} offered the most comprehensive review of MPC for buildings to date, including a unified framework that brings together theoretical foundations, practical implementation strategies, and performance assessment methods to close the gap between research and real-world applications. Another derailed review was given by Yao and Shekar \cite{Yao2021-lz}, classifying MPC types, addressing plant-model mismatches and disturbances, and highlighting optimization and implementation strategies. A recent review by Taheri et al. \cite{Taheri2022-yc} addressed topics such as prediction horizons and constraints, looking at several MPC methods and future research directions.

Some topical reviews look at the targeted aspects of MPC applications, addressing specific challenges in building control. On commercial buildings, Hilliard et al. \cite{Hilliard2016-ng} examined trends and opportunities for MPC and highlighted its benefits while using building thermal mass to optimize energy use and occupant comfort. On a similar subject, Rockett and Hathway \cite{Rockett2017-at} analyzed the challenges of MPC implementation in commercial buildings, pointing out the importance of predictive modeling and sensor integration for energy and operational efficiency. On occupancy behavior based MPC, Mirakhorli and Dong \cite{Mirakhorli2016-wq} investigated the potential of MPC to improve energy efficiency and occupant comfort via advanced modeling, prediction techniques, and field studies. Afram et al. \cite{Afram2017-ss} showed that using ANN-based MPC could help address non-linear HVAC dynamics and improve energy efficiency, as demonstrated through a residential case study. Thieblemont et al. \cite{Thieblemont2017-fd} reviewed MPC applications for thermal energy storage in buildings and highlighted how weather forecasts were used to optimize heating and cooling operations. Lastly, Tarragona et al. \cite{Tarragona2021-mf} investigated the role of MPC in renewable energy sources, in addition to building-level controls. They also stressed the use of MPC in managing renewable energy variability and operating hybrid energy storage systems.

Some researchers also reviewed MPC within broader frameworks of building controls. Benndorf et al. \cite{Benndorf2018-dj} incorporated MPC in their review of energy performance optimization in buildings and discussed its potential within a framework that deals with semantic interoperability and fault detection. Similarly, Mariano-Hernandez et al. \cite{Mariano-Hernandez2021-jw} reviewed energy management strategies, in particular MPC and static optimization, for demand-side management and fault detection. Kathirgamanathan et al. \cite{Kathirgamanathan2021-gk} reviewed the role of data-driven predictive controls in unlocking building energy flexibility through advanced modeling, grid integration, and practical applications. While not strictly a review, Stoffel et al. \cite{Stoffel2023-sw} assessed energy savings, comfort, and computational effort through simulations and compared MPC variants, including white-box, gray-box, black-box, and approximate MPC, to RL. Xin et al. \cite{Xin2024-gb} focused on model-based and model-free predictive controls and highlighted their applications in cooling, heating, and integrated systems while highlighting strengths, limitations, and advancements in methods like RL.

\subsection{Reviews of RL for buildings} %%%%%%%%%% 13 papers

In response to the growing adoption of RL in various fields like video games, targeted advertising, robotics, and autonomous vehicles, there has been a marked growth in studies investigating its application in buildings, especially since 2017. The earliest known review on this topic is by Vázquez-Canteli and Nagy \cite{Vazquez-Canteli2019-ey}, who reviewed 105 studies on RL algorithms in the context of demand response for smart grids. They emphasized the lack of real-world implementations in the literature and the difficulty of comparing results across different studies due to insufficient standardization of methods, simulation tools, and system dynamics. Mason and Grijalva \cite{Mason2019-jh} reviewed RL applications in building energy management systems. They found that the specific RL application significantly impacted overall energy savings, with savings of around 10\% for HVAC, 20\% for water heaters, and over 20\% for whole-building energy management. The authors reiterated that most studies were conducted only in simulations. Wang and Hong \cite{Wang2020-lx} performed a more in-depth review of the formulation of an RL controller for buildings, structuring their review into five key aspects of RL: algorithms, states, actions, rewards, and environment. They found that 77\% of the reviewed papers used value-based algorithms like Q-Learning, 91\% did not consider historical states, potentially violating the Markovian assumptions underlying RL theory and leading to suboptimal solutions. Also, only about 11\% of the 77 studies were conducted in real buildings. Fu et al. \cite{Fu2022-mq} classified different families of RL algorithms and analyzed their suitability to solve different problems related to efficient building control. A recent introductory review by Nagy et al. \cite{Nagy2023-ke} summarized some of the key modern trends in RL research for building energy management in a ten-question format, with an objective to guide its practical adoption and highlight potential future research directions. Critical highlights were the limited number of experimental studies, as well as the lack of safety and stability guarantees that some model-based control frameworks can provide. Unlike traditional reviews that only focus on RL algorithms previously implemented in other studies, Al Sayed et al. \cite{Al-Sayed2024-etal} sought to provide a link between theoretical RL concepts and the existing gap present in studies for their application in buildings. They concluded that novel techniques such as meta-RL are needed for the periodic extensive retraining of RL agents.

In addition to introductory and comprehensive reviews, several authors published topical reviews and those that discussed RL within broader frameworks of building controls. Kathirgamanathan et al. \cite{Kathirgamanathan2021-gk}, broadly reviewed data-driven control methodologies, including RL and MPC. With a particular focus on model-free HVAC control, Michailidis et al. \cite{Michailidis2023-gs} presented recent advancements in RL, ANNs, fuzzy logic control, and their hybrids. Sierla et al. \cite{Sierla2022-ac} categorized RL studies based on the decision of action modeling and analyzed the impact of specific choices, such as discrete, binary, or continuous action modeling. In a review by Shaqour et al. \cite{Shaqour2022-fh}, the authors examined the subset of studies that used deep reinforcement learning for various building types, by also discussing research directions and knowledge gaps. These modeling reviews are akin to the recent review of Michailidis et al. \cite{Michailidis2024-vv} that surveyed all distributed multi-agent-based building control studies. On the topic of occupant comfort, Han et al. \cite{Han2019-pf} reviewed RL applications and highlighted its under-use, in particular for indoor air quality and lightning, and the need for more OCC approaches. On a similar topic, Chatterjee and Khovalyg \cite{Chatterjee2023-yn} recently discussed the potential of RL to enable dynamic indoor environments that balance energy efficiency and comfort.

\subsection{Reviews of MPC and RL field demonstrations}

While many review papers discuss the potential of using MPC and RL in building HVAC control, few specifically focus on field demonstrations. In a partial review of MPC field demonstrations in commercial buildings, Sturzenegger et al. \cite{Sturzenegger2016-vd} categorized 10 papers based on the controlled portion of the building, actuator type, experiment duration, and underlying thermal model. Their review revealed gaps in implementation and cost considerations in the existing literature. Alongside their review, the authors performed a seven-month MPC experiment in a Swiss office building and demonstrated that MPC provides significant energy savings and comfort improvements; however, it requires high implementation costs. Blum et al. \cite{Blum2022-dc} investigated 17 papers on MPC field demonstration in commercial buildings and provided details on the key aspects of each paper, such as building type, system type, control variables and objectives, MPC approaches, test periods, and results. They also developed and tested an MPC system using a Modelica-based \cite{Wetter2014-dp} workflow in a real office building, proving its feasibility and addressing practical implementation challenges. In addition, they evaluated the implementation effort in person-days and identified barriers involved with integration, data management, and system commissioning. Narrowing their focus to residential cases, Pergantis et al. \cite{Pergantis2024a} focused on 12 relevant papers on field demonstration of MPC, covering their control methods, locations, equipments, control actions, test durations, and performance results. Pergantis et al. discussed key research gaps, such as limited information on deployment costs and a narrow focus on energy cost reduction. They deployed their MPC controller in an occupied, all-electric house in a cold climate and discussed implementation challenges. The most comprehensive review to date is that of Khabbazi et al. \cite{Khabbazi2024-im}, with a focus on 56 field tests in commercial buildings documented in 36 papers between 2005 and 2024. This study covered key aspects of papers such as control methods, building types, objectives, test durations, and reported results. Khabbazi et al. also introduced duration-weighted averages to assess reported savings and consequently showed the ability of MPC and RL to save energy and costs compared to conventional controllers, with lower savings demonstrated in longer-duration and whole-building tests. The authors highlighted persistent challenges, such as limited deployment data and scalability issues. In a recent work \cite{Saloux2025-yt}, Saloux conducted a partial review focused mainly on MPC and looked at methods for estimating energy savings from field demonstrations. Saloux analyzed about 40 papers on residential and commercial buildings and explored how different benchmarking methods affect reported savings. According to the findings, there is significant variability in savings estimates depending on the benchmarking method, which emphasizes the need for a more consistent assessment of field-demonstrated energy conservation measures.

\section{Methodology}

\label{sec:methodology}

\subsection{Search strategy}

This paper's literature review began by identifying studies that used MPC or RL in real-world residential or commercial buildings. Peer-reviewed publications, including journal articles and refereed conference proceedings, were found using search engines such as Google Scholar and Scopus, as well as databases from publishers such as Elsevier, IEEE, Taylor \& Francis, Springer and Wiley. When a more detailed journal version of a conference paper was available, the conference version was excluded. The figures and calculations in this paper do not include theses, dissertations, white papers, or technical reports.

To manage the large body of literature, specific search strategies were used. Keywords such as ``field demonstration,'' ``experimental study,'' ``residential building,'' ``commercial building,'' ``model predictive control,'' ``reinforcement learning,'' and ``HVAC'' were frequently used to refine the search and identify relevant papers. These are only a few of the many keywords used during this process. The identified publications were organized into four tables (residential vs. commercial, papers vs. tests), each with a wide range of study characteristics as separate columns. A full, open-access version of this table can be found in \cite{field-demonstrations-HVAC-MPC-RL}. \ref{sec:appendixa} contains two more concise versions of these tables, summarizing 24 residential (\autoref{tab:res}) and 80 commercial (\autoref{tab:com}) papers, respectively. These tables include details such as publication year, control method, building and controlled space description, HVAC system description, test durations, objectives, and reported benefits along with the benchmarking approach.

During the review process, we often contacted authors to gather information that we could not find in the reviewed papers. We contacted authors via email, LinkedIn, and in-person meetings at workshops and conferences. Of the 104 field studies reviewed, 45 required outreach to authors for further clarification. We reached out to 32 unique authors (some who wrote multiple papers), receiving responses from 26, with an 81\% response rate. A full list of the responding researchers is provided in the \nameref{sec:acknowledgments}. Questions about the other six papers remained unresolved due to, for example, changes in author affiliations or lack of contact information. For these studies, we tried to collect missing data by reading related conference papers, technical reports, theses, and dissertations. We also attempted to estimate parameters such as floor areas and locations ourselves using tools such as Google Maps.

\subsection{Savings evaluation}

Occasionally, a reviewed paper presented findings from multiple field tests conducted in multiple building types, regions, and climate zones; or using multiple control methods. When reviewing key results, such as reported benefits, this paper considers the total number of field tests rather than the number of papers. The reported performance benefits in the literature include energy and cost savings, comfort improvement, greenhouse gas emissions reduction, peak demand reduction, frequency regulation improvement, and more. The main focus of this paper is on the energy and cost savings achieved by using MPC and RL over conventional control methods. Since longer-duration tests tend to be more reliable than short-duration ones, this paper introduces the duration-weighted average percent savings shown in Eq. \hyperref[eq:savings]{\ref{eq:savings}}:
\begin{equation}
\overline{\% \text{ savings}} = \frac{\sum_{i} (\% \text{ savings})_i \cdot (\text{duration})_i}{\sum_{i} (\text{duration})_i} .\label{eq:savings}
\end{equation}
Here, $\overline{\mathrm{\% \text{ savings}}}$ denotes the duration-weighted average percent savings. The integer $i$ indexes field tests. This approach gives more weight to tests with longer durations.

\subsection{Statistical methods}

This paper uses the Mann-Whitney U test \cite{mann1947test} to compare categorical variables in two groups, such as experiments that controlled a whole-building vs. a subset or building spaces, that used measurement-based vs. simulation-based benchmarks, and that used test durations of less than seven days vs. at least seven days. The Mann-Whitney U test is a nonparametric test of the null hypothesis $H_0$ that the populations underlying both categories have the same distribution, vs. the alternative hypothesis $H_1$ that the populations have different distributions. We chose the Mann-Whitney U test in order to handle data that may not follow Gaussian distributions. Statisticians typically recommend a minimum sample size of 15 per category to guarantee trustworthy results. Consequently, we do not report test results for categories having fewer than 15 samples. Due to these constraints, some comparisons could not be tested rigorously.

\section{Demographics}

\label{sec:demographics}

\begin{figure*}[!t]
    \centering
    \includegraphics[width=0.975\textwidth]{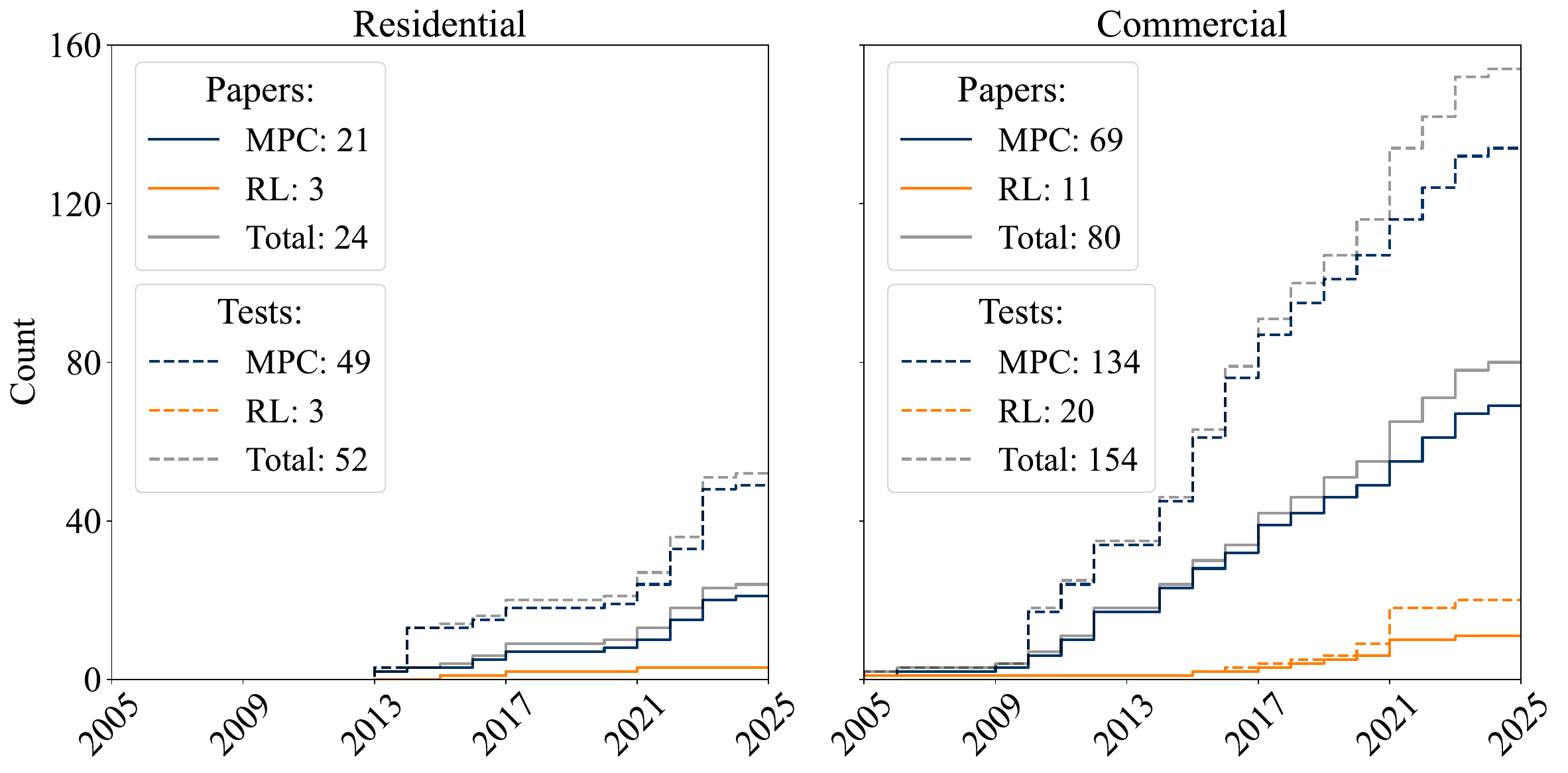}
    \caption{Cumulative counts of residential (left) and commercial (right) field demonstrations of MPC and RL, categorized by papers (solid) and tests (dashed).}
    \label{FIG2}
\end{figure*}

Using the methodology outlined above, the final search identified 24 residential papers (52 field tests) and 80 commercial papers (154 field tests). \autoref{FIG2} shows the cumulative counts from 2005 to 2025 of papers (solid curves) and field tests (dashed) of MPC and RL in residential (left) and commercial (right) buildings. The earliest field demonstration of MPC in a commercial building was conducted in 2005 at the Energy Resource Station in Ankeny, Iowa, USA \cite{Henze2005-ed}. Since then, the count of MPC publications in commercial buildings has steadily grown, resulting in 69 papers covering a total of 134 field tests. The first residential MPC field demonstration was carried out in 2013 using single-family detached homes in Denmark \cite{Pedersen2013-ho}. Following this effort, residential MPC research has grown consistently, with 21 papers reporting 49 field tests to date. RL saw its first application in a commercial building in 2006 at the Energy Resource Station \cite{Liu2006-ik}. No further studies appeared until 2016, when research gradually increased to the current totals of 11 papers and 20 field tests. The first residential RL field demonstration was conducted in a test room in Leuven, Belgium \cite{Leurs2016-mq}. Only two residential RL papers have been published since then, each reporting a single five-day field test. 

\subsection{Geography}

\begin{figure}[!t]
    \centering
    \includegraphics[width=0.975\textwidth]{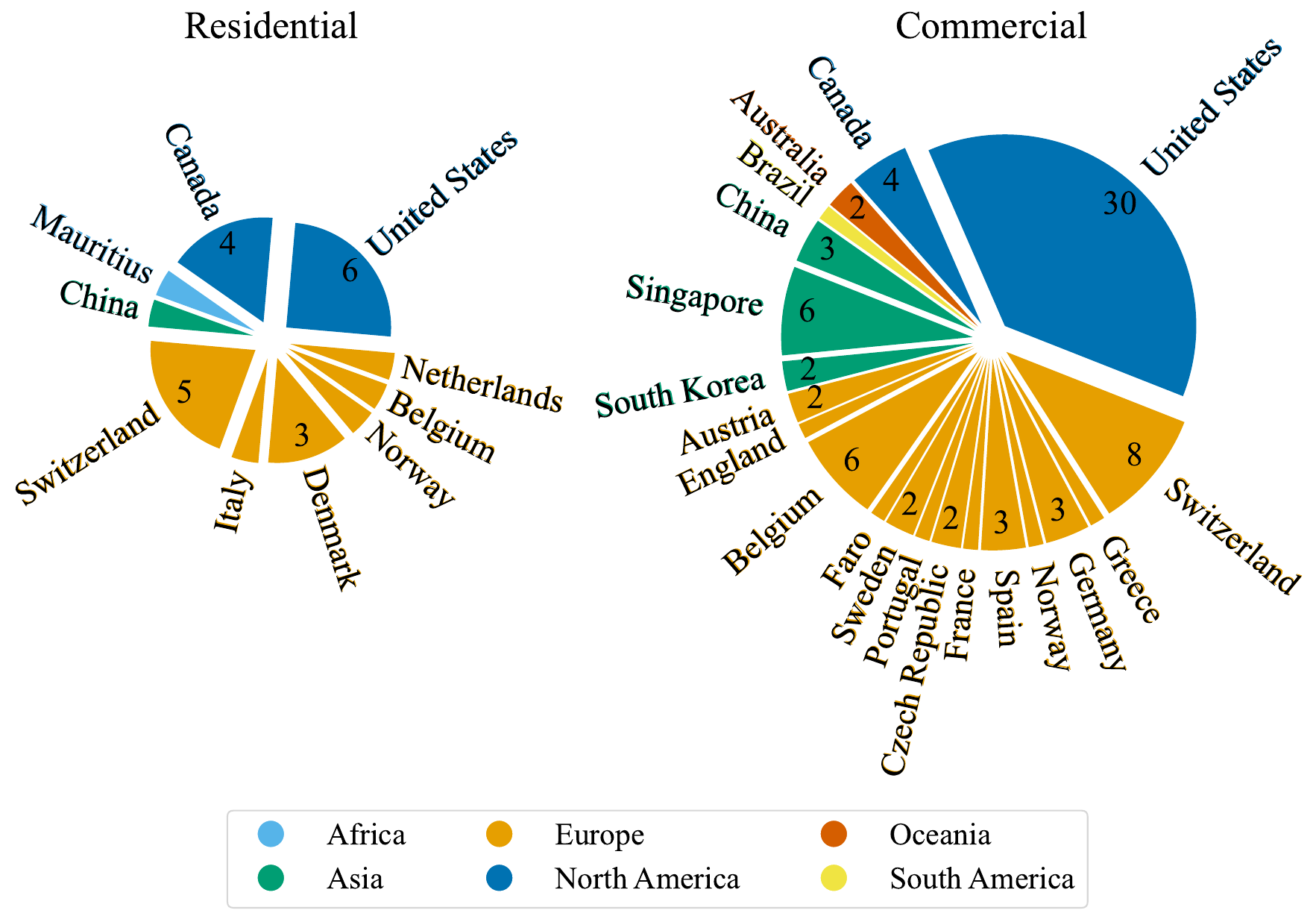}
    \caption{Geographical distribution of residential (left) and commercial (right) field demonstration papers by country and continent. Colors indicate continents. Chart sizes are proportional to total numbers of papers. Slices without numbers correspond to a single paper.}
    \label{FIG3}
\end{figure}

\autoref{FIG3} shows the distribution of papers by location, classified by country and continent, for residential (left) and commercial (right) buildings. The size of each pie chart corresponds to the total number of papers in its category. Each continent is uniquely color-coded. Both pie charts show that the bulk of papers originated from North America and Europe, with the United States and Switzerland leading their respective continents with around 35\% and 13\% of all papers, respectively. While some continents are classified in a single category, the combined total of residential and commercial papers includes contributions from all continents. Despite having over half of the world's population, Asia accounts for only 12 of the 104 papers (12\%). Other regions have much lower representation, with only one paper from Africa, one from South America, and two from Oceania.

\subsection{Weather}
\label{weather}

\begin{figure}[!t]
    \centering
    \includegraphics[width=0.975\textwidth]{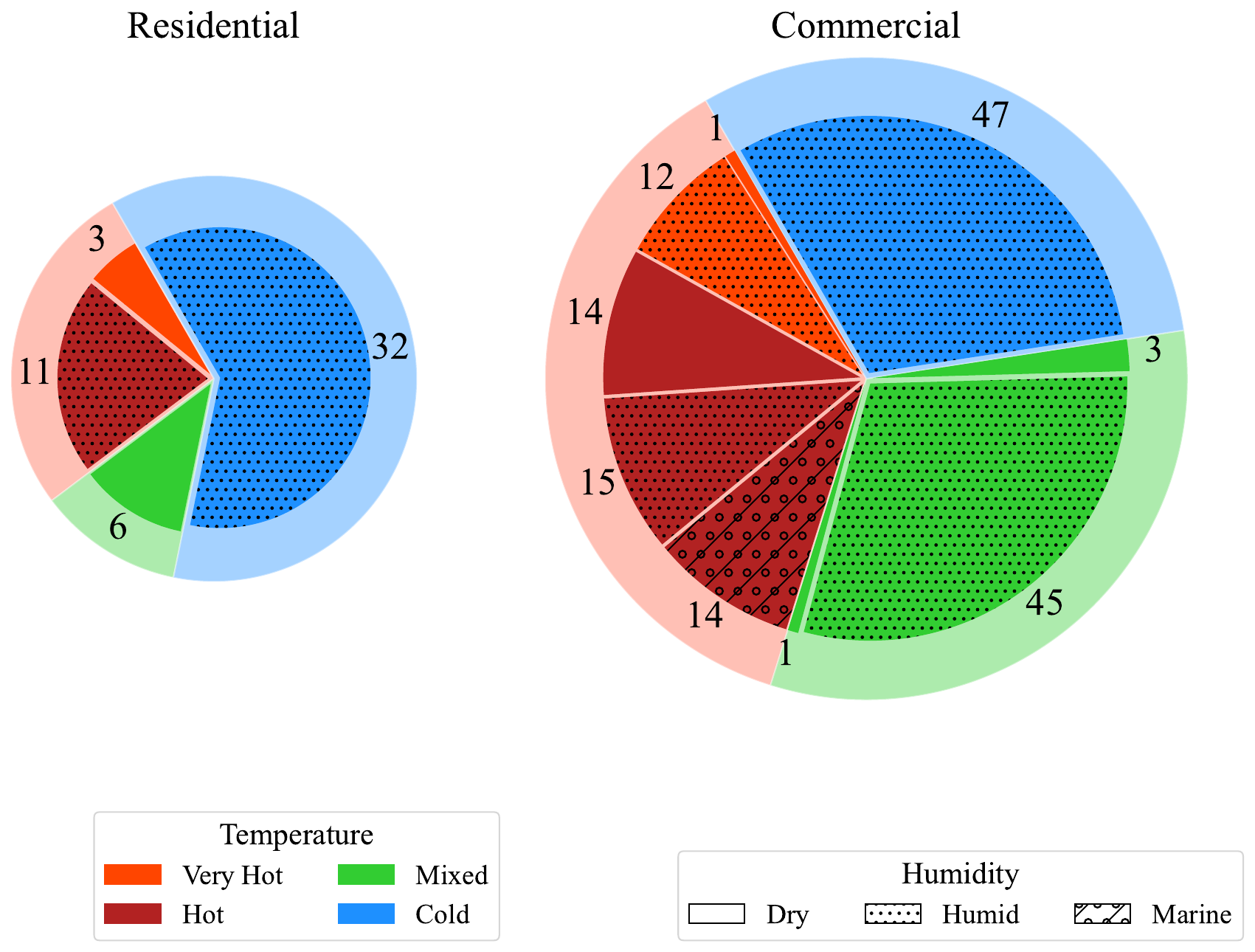}
    \caption{Distribution of residential (left) and commercial (right) field tests by climate zone. Colors indicate temperature zones; hatching patterns indicate humidity levels. Chart areas are proportional to total numbers of tests.}
    \label{FIG4}
\end{figure}

\autoref{FIG4} shows the distributions of residential (left) and commercial (right) field tests by climate zone. ASHRAE Standard 90.1 \cite{ASHRAE2022} defines climate zones based on temperature and humidity characteristics, which provides a framework for categorizing regions. The distributions in \autoref{FIG4} refer to the number of tests performed, not the number of papers, as opposed to \autoref{FIG3}. This difference is because some papers reported multiple tests carried out in different regions. The size of each chart corresponds to the total number of tests in that group. Colors also indicate temperature zones and hatching patterns the humidity levels.

Most residential tests (32) were performed in cold climate zones, followed by hot (11), mixed (6), and very hot (3). All 52 residential tests were carried out in humid regions, with no tests recorded in marine or dry locations. Commercial cases have a more even distribution across temperature zones, with 49 tests in mixed, 47 in cold, 43 in hot, and 13 in very hot zones. An additional exterior shaded pie chart divides the temperature zones into three groups: red (extremely hot and hot), green (mixed) and blue (cold). These groups have nearly equal distributions, with 56, 49, and 47 tests, respectively. In terms of humidity, most of the commercial tests (119) took place in humid areas, with fewer tests recorded in the marine (17) and dry (16) regions.

\subsection{Experiment scope} 

\begin{figure*}[!t]
    \centering
    \includegraphics[width=0.975\textwidth]{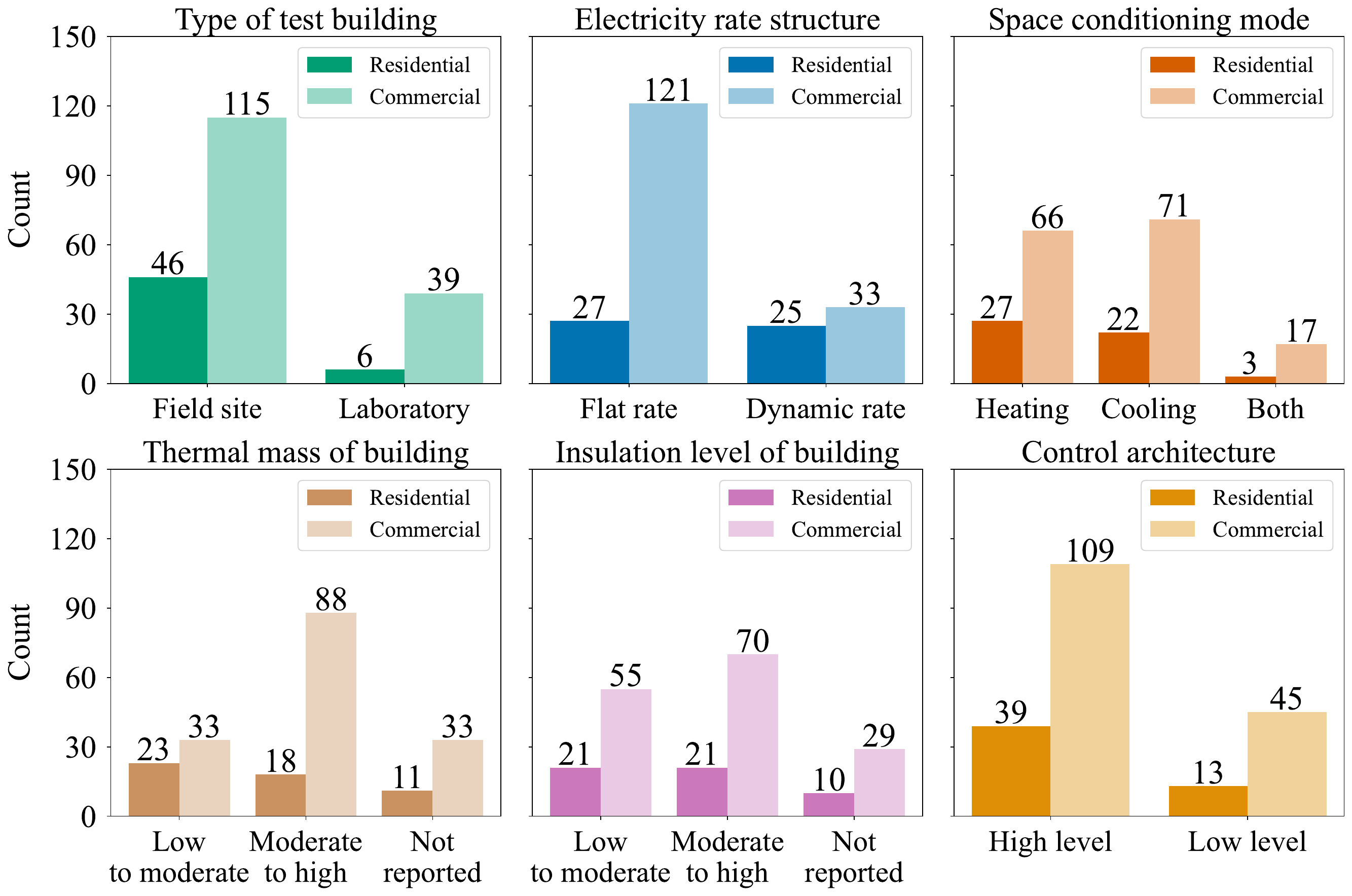}
    \caption{Numbers of residential and commercial field tests within each category of experiment scope.}
    \label{FIG5}
\end{figure*}

Several key characteristics define field experiment scopes. These include the type of test building (e.g., an occupied field site vs. a laboratory environment), the control architecture (e.g., supervisory setpoint adjustments vs. direct actuation of HVAC components), the electricity rate structure (e.g., time-invariant price vs. time-of-use or day-ahead prices), the space conditioning mode(s) considered (e.g., heating, cooling, or both), and the properties of the building, such as its insulation and thermal mass. \autoref{FIG5} summarizes these characteristics for experiments conducted in residential (darker bars) and commercial (lighter) buildings.

\subsubsection{Type of test building}

Experimental demonstrations of MPC and RL may be carried out in field sites or laboratories. Field sites include operational settings such as offices, university buildings, hospitals, schools, houses, apartments, and other comparable facilities. In contrast, laboratories are controlled environments, such as test chambers and research stations, that typically are unoccupied and see little use beyond research. Field sites offer more realistic test conditions, while laboratories offer greater reproducibility. The top left plot of \autoref{FIG5} shows that 46 of 52 (88\%) residential tests were conducted at field sites. Similarly, 115 of 154 (75\%) commercial tests were conducted at field sites. Six residential tests and 39 commercial tests were conducted in laboratories.

\subsubsection{Electricity rate structure}
\label{ratestructure}

Electricity rate structures determine how energy prices vary, ranging from a flat rate (meaning a single, time-invariant price) to dynamic rates (meaning prices that vary with time) such as time-of-use or day-ahead pricing. Electricity rate structures influence the potential for cost savings by load shifting. The top center plot in \autoref{FIG5} shows the distribution of tests by electricity rate structure. Residential tests are almost equally balanced between flat (27 of 52 tests, or 52\%) and dynamic rate structures (25 of 52 tests, or 48\%). In contrast, commercial tests strongly favor flat rates, with 121 of 154 tests (79\%) using flat rates and 33 (21\%) using dynamic rates.

\subsubsection{Space conditioning mode(s)}
\label{operationalmode}

Researchers have evaluated MPC and RL for HVAC systems operating in heating mode, cooling mode, or both. Residential field demonstrations that evaluate both heating and cooling typically span several months to include both cold and hot weather. In some commercial buildings, however, HVAC systems may simultaneously heat some thermal zones and cool others. In such commercial cases, it was impossible to distinguish between heating and cooling effects because studies reported only one aggregate value rather than separate values for each operational mode. The top right plot of \autoref{FIG5} shows the distribution of tests of heating, cooling, or both. Heating and cooling modes are distributed nearly evenly in both residential and commercial cases, with fewer tests incorporating both modes. In the residential category, 27 (52\%) of 52 tests were performed in the heating mode, 22 (42\%) in cooling mode, and only three (6\%) in both modes. For commercial buildings, 66 (43\%) of 154 tests focused on heating, 71 (46\%) on cooling, and 17 (11\%) on both.

\subsubsection{Thermal mass of building}
\label{thermalmass}

Building thermal mass affects temperature stability and load-shifting effectiveness by influencing the building's ability to store thermal energy \cite{Raman2020-lm}. Early studies in the twenty-first century explored the role of thermal mass, such as optimizing active and passive thermal storage under predictive control \cite{Henze2004-jy} and evaluating thermal mass control strategies for cost savings and peak demand reduction \cite{Braun2001-kd}. Buildings with high thermal mass smooth out temperature fluctuations and reduce sensitivity to short-term disturbances \cite{Kavgic2015-eo}. Additionally, predictive control that activates thermal mass can improve energy efficiency and comfort \cite{Wang2023-ow}. The lower left plot of \autoref{FIG5} categorizes studies based on thermal mass into low to moderate, moderate to high, or not reported. Due to a lack of standardized reporting, we classify tests based on available descriptions of the building envelope within the paper itself, previous studies on the same building, relevant institutional or project websites, author correspondence, and, in rare cases, reverse image searches of the building. If none of these methods yielded a reasonable estimate, the thermal mass was marked ``not reported.'' Despite these efforts, almost one-fifth of residential (11 of 52) and commercial (33 of 154) tests lacked adequate thermal mass information. The reported cases for residential buildings were almost evenly distributed, with 18 tests rated moderate to high thermal mass and 23 as low to moderate. Commercial buildings, however, showed a stronger preference for higher thermal masses, with 88 tests classified as moderate to high and only 33 as low to moderate. This suggests that commercial structures are more often constructed with heavier materials such as brick and cement, whereas lighter wood-frame construction, while present in both sectors, is more common in residential buildings.

\subsubsection{Insulation level of building}
\label{insulation}

Building insulation increases thermal energy retention and can improve load shifting potential \cite{KimBraun2022}. The lower middle plot of \autoref{FIG5} shows the classification of studies based on insulation levels: low to moderate, moderate to high, or not reported. Given the lack of standardized reporting, we classified tests using the same approach as that used for thermal mass assessment. Despite these efforts, almost one-fifth of residential (10 of 52) and commercial (29 of 154) tests did not provide insulation information. The reported cases were nearly evenly distributed, with 70 commercial and 21 residential tests classified as moderate to high insulation and 55 commercial and 21 residential tests classified as low to moderate.

\subsubsection{Control architecture}
\label{control}

HVAC control architectures can be classified as either high-level or low-level. High-level control typically entails supervisory adjustment of setpoints, such as zone air temperatures or supply air temperatures, with device-level controllers such as proportional-integral loops to track high-level setpoints. Low-level control, however, directly actuates components such as dampers, valves, and compressors, augmenting or replacing the manufacturer's device-level control logic. The lower right plot of \autoref{FIG5} shows the distribution of tests classified by their control architecture. High-level control was used in 39 (75\%) of 52 residential tests and 109 (71\%) of commercial tests, while low-level control was used in 13 (25\%) and 45 (29\%), respectively.

\subsection{Algorithm implementation choices}

This section summarizes the specific algorithm implementation choices that most researchers made in the field papers reviewed here. For tutorial reviews of the MPC and RL algorithms and implementation options in the context of HVAC, we refer the reader to \cite{Serale2018-ph, Drgona2020-fh, Yao2021-lz} for MPC and \cite{Vazquez-Canteli2019-ey, Wang2020-lx} for RL.

\subsubsection{MPC implementation}

The basic components of an MPC algorithm are the system model; a method for forecasting uncertain model inputs; an optimization model including decision variables, an objective function, and constraint functions; and a numerical optimization solver. When state variables are imperfectly observed, MPC is typically combined with a recursive state estimator, such as a Kalman filter for linear systems, or for nonlinear systems, a particle filter or an extended or unscented Kalman filter.

% system model
Two-thirds of the MPC field papers used linear models of the building dynamics. Most models followed a thermal circuit structure (also known as an RC network), with a typical model order of one to three states per thermal zone. Linear autoregressive time-series models were also fairly common. Smaller numbers of papers modeled system dynamics using high-dimensional, nonlinear building simulation software such as EnergyPlus \cite{crawley2001energyplus}, TRNSYS \cite{klein1976trnsys}, or Modelica. Other papers used nonlinear regression models, such as neural networks or regression trees. Most papers used time step durations of five to 60 minutes. HVAC equipment performance curves were typically modeled using polynomials of degree one to three, often fit to manufacturer performance data, although a few papers used lookup tables or neural networks.

% forecasting
The MPC field papers used a wide range of methods for forecasting uncertain model inputs such as internal heat gains from sunlight, electrical loads, and bodies. Popular forecasting methods include linear autoregressive time-series models and neural networks. In addition to implementing their own forecasting methods, many papers also downloaded external forecasts of weather variables or energy prices. Most papers used forecast horizons of two to 24 hours.

% optimization problem
The MPC field papers often minimized the cumulative input energy to HVAC equipment over the forecast horizon or the cumulative energy cost. A small number of papers aimed, additionally or instead, to minimize the peak electricity demand over the forecast horizon, or to maximize revenues from providing demand response or power grid reliability services. Many papers augmented the objective function with `soft constraints,' meaning constraint violation penalties. Indoor air temperature constraints were often `softened' to avoid infeasible optimization problem instances. Most papers enforced HVAC equipment limits via (hard) constraints, rather than (soft) penalties. Indoor temperatures and heat transfer rates were common optimization variables, although some papers directly optimized equipment set-points, such as air or water flow rates, supply temperatures, compressor speeds, or fan speeds.

% solver details
After combining dynamical models, forecasts, and optimization models, the MPC field papers arrived at a wide range of optimization problem structures. Many papers solved convex optimization problems -- typically linear or quadratic programs -- for which off-the-shelf solvers can return globally optimal solutions in polynomial time \cite{boyd2004convex}. Many other papers solved nonconvex optimization problems, with nonconvexities typically arising from nonlinear dynamics, nonconvex equipment performance curves, or integer-valued decision variables representing, for example, equipment on/off states. Most nonconvex optimization solvers cannot guarantee global solution in polynomial time, but often return locally optimal solutions that significantly improve upon user-specified initial guesses.

\subsubsection{RL implementation}

The basic components of an RL algorithm are the state and action spaces, reward function, and exploration method. In HVAC applications, deploying an untrained RL agent in hardware could put equipment at risk or require long training times before convergence to an effective policy. For these reasons, all RL field papers reviewed here pre-trained an agent offline in a simulator before deploying it in hardware. Implemented in this way, even model-free RL variants require a system model for offline pre-training. The most common system models used in RL field papers were neural networks, followed by EnergyPlus and the low-order linear thermal circuits that are popular in MPC.

Nearly all RL field papers used the Q-learning or actor-critic RL variants. Typical actions included equipment on/off states, indoor air temperature set-points, supply fluid flow or temperature set-points, or fan speeds. Nearly all RL papers included indoor air temperatures as states; many also included states related to the time of day, weather measurements or forecasts, and occupancy measurements.

Like the objective function in MPC, the RL reward function encodes user-specified goals. As with MPC, most RL field papers aimed to reduce input energy or energy costs. Most RL variants do not readily accommodate constraints, so papers typically augmented the reward function with violation penalties associated with thermal comfort or equipment limits. Several papers also rewarded temporal smoothness of the action signal.

Exploration methods balance exploration/exploitation trade-offs by steering the system toward state-action configurations that may be under-explored. The most popular exploration methods in HVAC RL field papers were the $\epsilon$-greedy and Boltzmann approaches. A few papers encouraged exploration by adding Gaussian noise to the implemented actions.

\subsection{Experiment protocol choices} 

Section \ref{sec:protocols} discusses three experiment protocol choices that can influence the accuracy of energy and cost savings estimates. Section \ref{sub:benchmark} discusses the choice of benchmark against which to assess energy and cost savings (e.g., the measured performance of a baseline control system in the real building vs. the simulated performance of a baseline control system in a model of the building). Section \ref{sub:controlledportion} discusses the choice of the portion of the building controlled (e.g., the whole building vs. a subset of building spaces, such as a limited number of rooms, thermal zones, or floors). Section \ref{sub:duration} discusses the choice of experiment duration, which ranged from one day to one year in the papers reviewed here.

\begin{figure*}[!t]
    \centering
    \includegraphics[width=0.975\textwidth]{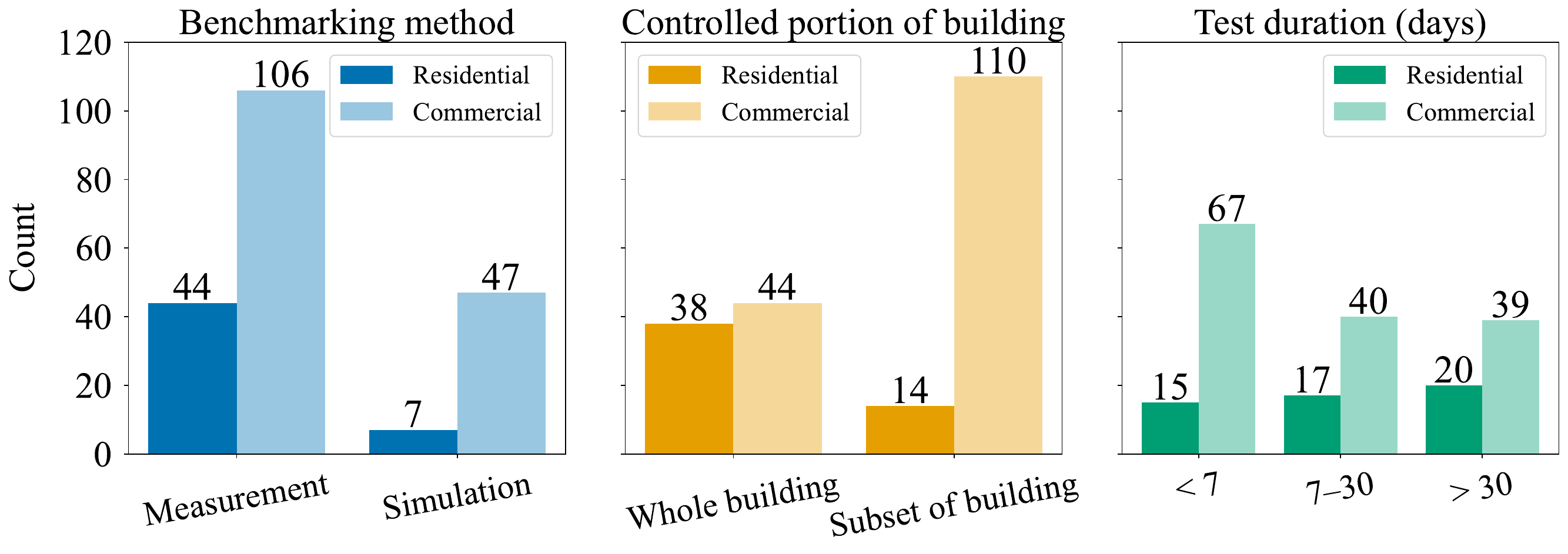}
    \caption{Numbers of residential and commercial tests within each category of experiment protocol choices.}
    \label{FIG6}
\end{figure*}

\autoref{FIG6} shows the numbers of residential (darker bars) and commercial (lighter) field tests corresponding to each choice in the three experiment protocols. The left plot shows that the majority of both residential and commercial tests benchmarked against measurements. However, seven (14\%) of 51 residential tests (one not benchmarked) and 47 (33\%) of 153 (one not benchmarked) commercial tests benchmarked against simulations. The center plot shows that 38 (73\%) of 52 residential tests controlled a whole building, whereas 110 (71\%) of 154 commercial tests controlled a subset of building spaces. The right-most plot in \autoref{FIG6} shows that 15 (29\%) of 52 residential tests and 67 (46\%) of 146 (eight did not report) commercial tests lasted less than seven days. Only 20 (38\%) of 52 residential tests and 39 (27\%) of 146 commercial tests lasted one month or more.

\section{Experiment protocols}
\label{sec:protocols}

Experiment protocols varied significantly over the 104 papers reviewed here. Some field tests benchmarked energy and cost savings against measurement-based counterfactuals, while others used simulations. Similarly, some researchers controlled an entire building, while others controlled a subset of building spaces, such as a room, thermal zone, or floor. Test durations varied significantly, ranging from one day to one year. This section investigates the influence of experiment protocols on estimated energy and cost savings.

\subsection{Benchmarking method}
\label{sub:benchmark}

\begin{figure}[!t]
    \centering
    \includegraphics[width=0.975\textwidth]{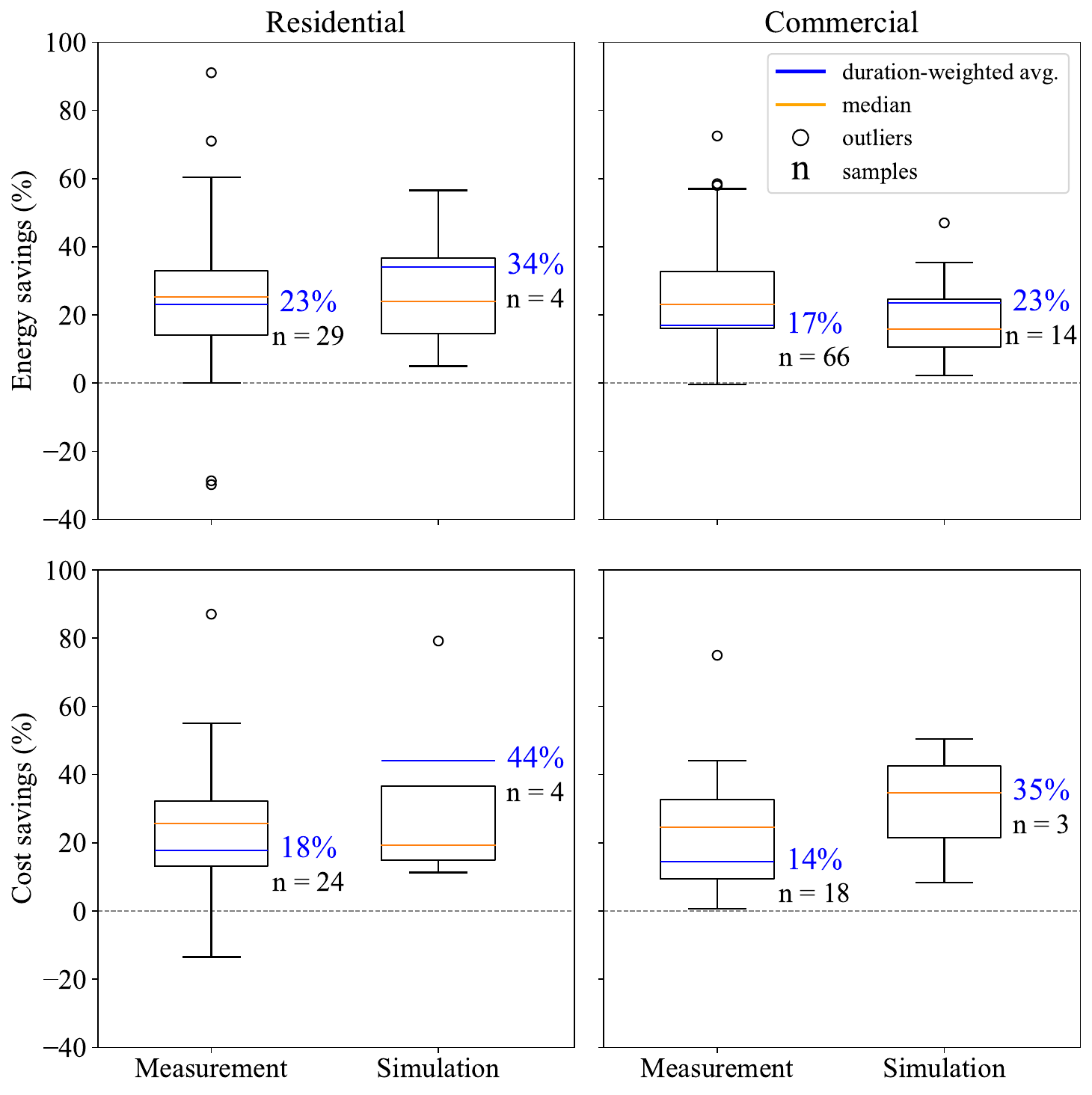}
    \caption{Energy and cost savings in residential and commercial tests, categorized by benchmarking method. Duration-weighted averages (blue) suggest that benchmarking against simulations systematically overestimates savings.}
    \label{FIG7}
\end{figure}

The choice of benchmark (also known as the baseline or counterfactual) influences the estimates of energy and cost savings from advanced HVAC control relative to a baseline control method. Ideally, the advanced control method and the baseline control method would be implemented in the same building under the same weather and occupancy conditions. However, this is only possible if researchers have access to two identical, side-by-side buildings, or to a building inside an environmental chamber where weather conditions can be repeatably emulated. Such experimental testbeds are nearly nonexistent, so researchers almost always must resort to less rigorous benchmarking methods, of which there are two broad categories. First, measurement-based benchmarks measure the field performance of the advanced control method and the baseline control method in the same building at different times, and therefore under potentially different weather and occupancy conditions. This approach typically involves adjusting for weather variations, for example, through normalization of energy use by heating degree-days or cooling degree-days. Second, simulation-based benchmarks compare the measured performance of the advanced control method in the real building to the simulated performance of the baseline control method in a computer model of the building, typically under the same weather conditions observed in the advanced control test period. Common model structures include thermal networks, time-series models, ANNs, regression trees, support vector machines, and high-fidelity building energy simulation programs such as EnergyPlus and TRNSYS.

\autoref{FIG7} shows the energy (top row) and cost (bottom) savings reported by field tests in residential (left column) and commercial (right) buildings, broken out by benchmarking method (measurement-based vs. simulation-based). In these box-and-whisker plots, the orange lines are medians, the blue lines are duration-weighted averages (see Eq. \ref{eq:savings}), the boxes span the 25th to 75th percentiles, the whiskers span the 2.5th to 97.5th percentiles, and the circles are outliers. The results show a noticeable difference: Compared to measurement-based benchmarks, simulation-based benchmarks consistently show higher savings distributions. In particular, the duration-weighted average savings estimates -- the statistics that the authors view as the most meaningful -- are substantially higher for simulation-based benchmarks in all four quadrants of \autoref{FIG7}. This trend suggests that simulation-based benchmarks may systematically overestimate savings. Statistical analysis was considered to test the hypothesis that simulation-based benchmarks systematically overestimate savings relative to measurement-based benchmarks. However, due to the small sample sizes for simulation-based tests (all below the minimum threshold of 15 for robust statistical inference), no meaningful statistical conclusions could be drawn.

We conjecture that two effects could explain the systematic overestimation of savings when using simulation-based benchmarking. First, developing an accurate building model -- whether physics-based or data-driven -- is challenging. Models often have prediction error ranges that are comparable to the 10--20\% savings that advanced control typically delivers. Therefore, the signal-to-noise ratio is generally low when attempting to distinguish real savings (the signal) from prediction errors (the noise). Second, building models almost always have parameters or hyperparameters that researchers need to manually tune. This manual tuning process presents opportunities for researchers to `put a thumb on the scale,' inflating energy use or cost in simulations of baseline control to increase the estimated savings from advanced control. While we do not suggest that researchers would intentionally manipulate results in this way, we also recognize that researchers are under high pressure to publish, and that publishing more impressive results is generally easier. From the perspective of scientific ethics, we believe it is important to acknowledge this source of potential subconscious bias and to take measures to prevent it from influencing reported results.

Both simulation- and measurement-based benchmarking methods introduce uncertainty into energy and cost savings estimates. Reporting confidence intervals alongside point estimates can characterize the degree of uncertainty present, and thereby inform readers about the trustworthiness of savings estimates. However, confidence intervals are rarely reported in field experiments. Among residential papers, only three out of 23 (one not applicable) included confidence intervals, while just six out of 77 (three not applicable) commercial papers did the same. Overall, only 9\% of papers reported confidence intervals for their energy or cost savings estimates.

To promote reliability of energy and cost savings estimates, all subsequent analyses in this paper use only savings estimates that were calculated using measurement-based benchmarks. Simulation-based results are filtered out due to the risk that they overestimate savings. Filtering out simulation-based benchmarks removed 54 (26\%) of 206 tests from the original dataset.

\subsection{Controlled portion of building}
\label{sub:controlledportion}

\begin{figure}[!t]
    \centering
    \includegraphics[width=0.975\textwidth]{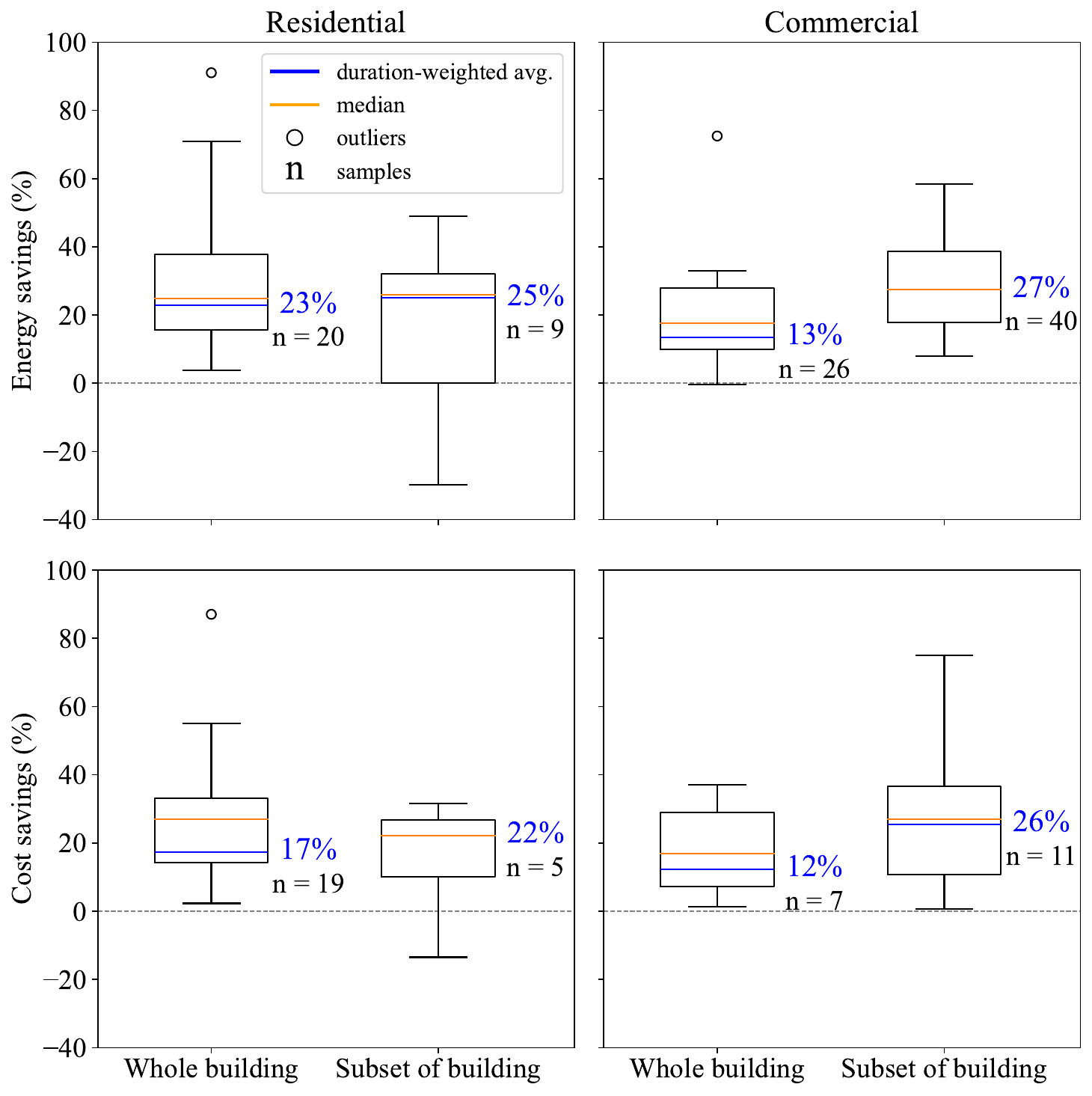}
    \caption{Energy and cost savings in residential and commercial studies, categorized by controlled portion of building. Controlling a subset of a building appears to systematically overestimate savings.}
    \label{FIG8}
\end{figure}

Field demonstrations may deploy advanced HVAC control systems in a whole building or in a subset of building spaces, such as a limited number of rooms, thermal zones, or floors. \autoref{FIG8} shows the distributions of energy and cost savings in residential and commercial buildings, classified by whether the control was applied to the whole building or a subset. Controlling a subset of a building appears to consistently result in higher savings estimates, especially in commercial buildings (which are usually much bigger than residential buildings). In commercial buildings, the duration-weighted average energy savings when controlling a subset of a building is 27\%, compared to 13\% when controlling a whole building. We tested the hypothesis that controlling a subset of a commercial building leads to systematically higher energy savings estimates using the Mann-Whitney U Test (U = 738, p = 0.004), which confirmed a statistically significant difference at the 1\% significance level. Cost savings followed a similar pattern, with subset-controlled cases saving 26\% vs. 12\% in whole-building control. However, statistical tests were not performed for cost savings due to the small sample sizes (7 for whole building cases and 11 for subset-controlled cases), which are less than the minimum threshold of 15 typically recommended for reliable statistical analysis.

In residential buildings, the difference is less pronounced between controlling a whole building vs. a subset of a building. The difference in duration-weighted averages is minimal, with subsets reporting slightly higher energy savings (25\%) than whole buildings (23\%). The differences in cost savings between the two categories for cost savings is more evident than those for energy savings, with 17\% reported for whole buildings and 22\% reported for subsets.

We conjecture that savings overestimation from controlling a subset of a building is the result of neglecting heat transfer between the controlled thermal zones and the adjacent zones. To illustrate the physical mechanism, we consider the simple example of controlling one zone in a two-zone building. In winter, an MPC or RL system might reduce the temperature of the controlled zone during unoccupied periods. This could result in high apparent savings when considering only the energy required to heat the controlled zone. However, reducing the temperature of the controlled zone would also cause heat transfer from the adjacent zone to the controlled zone. This heat transfer would increase heating demand in the adjacent zone. Reporting only the energy savings in the controlled zone -- and not accounting for the increased energy use to heat the adjacent zone -- would lead to overestimating the overall savings from MPC or RL. To avoid overestimating savings, researchers can control whole buildings or account for heat transfer with adjacent zones; we discuss this further in Section \ref{recommendations}.

To promote reliability of energy and cost savings estimates, subsequent analyses in this paper mainly use savings estimates from studies that controlled a whole building or a large subset of a building. Most studies that controlled a small subset of building spaces are filtered out due to the risk that they overestimate savings. However, we include papers \cite{Privara2011-xl, Siroky2011-tb, Vana2014-cu, Killian2018-nv, Drgona2020-es, Blum2022-dc, Lolli2024-hx} that controlled large subsets of commercial building spaces, leaving out only a few small zones due to restrictions imposed by building managers. Many of these studies also report good insulation between zones, suggesting negligible heat transfer with adjacent zones. In contrast, some whole-building studies controlled small test chambers; we exclude those studies from the final analysis to focus on real-world applications. Specifically, we exclude whole-building studies with floor areas below 100 m$^2$ and include subset-of-building studies with floor areas above 1000 m$^2$. These filters removed 70 (46\%) of the 152 tests that remained after filtering out studies that used simulation-based benchmarks.

\begin{figure}[!t]
    \centering
    \includegraphics[width=0.975\textwidth]{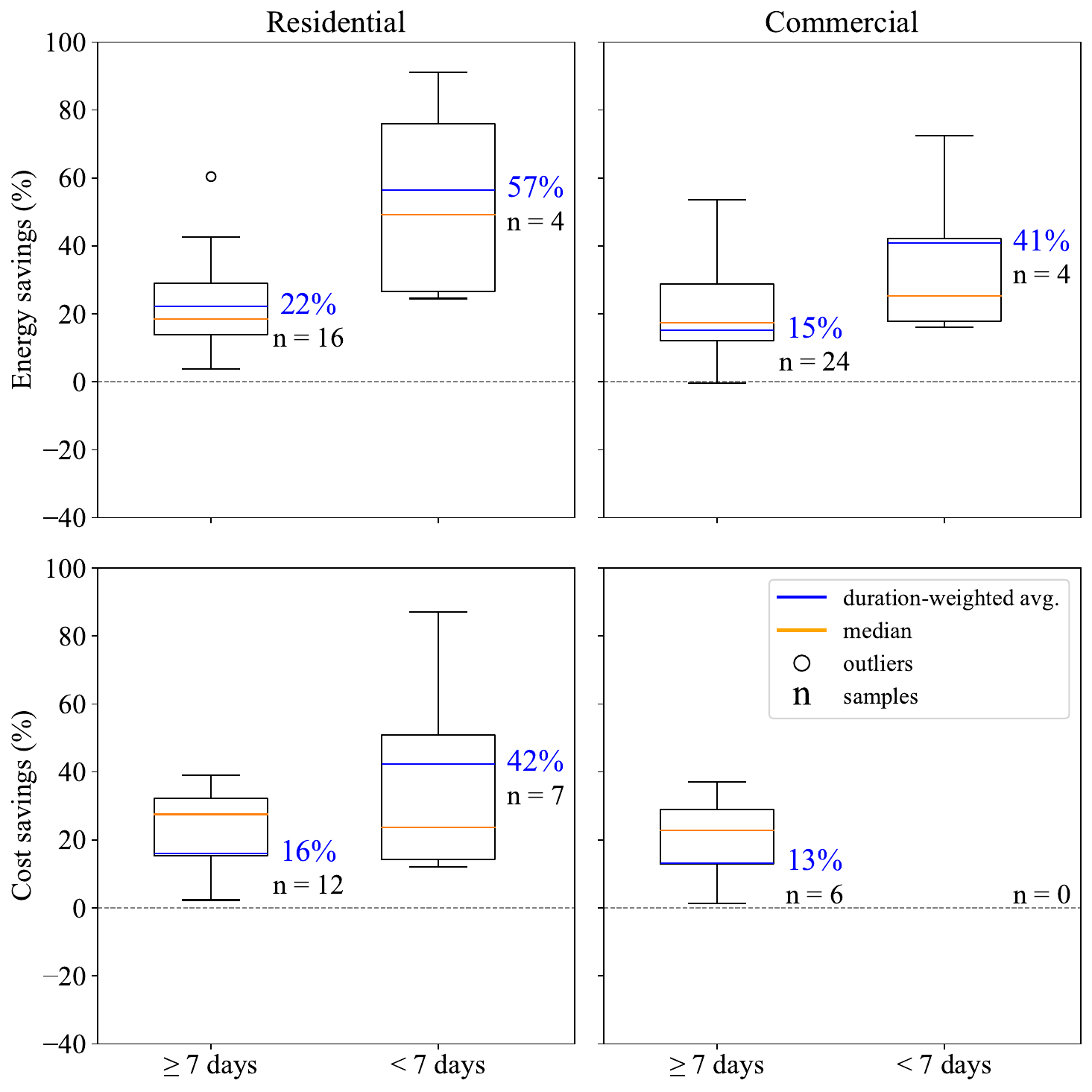}
    \caption{Energy and cost savings in residential and commercial studies, categorized by test duration. Testing for a short duration appears to systematically overestimate savings.}
    \label{FIG9}
\end{figure}

\subsection{Test duration}
\label{sub:duration}

The reviewed field demonstrations test advanced HVAC control for a wide variety of experiment durations, ranging from one day to one year in the unfiltered dataset. As one might expect, shorter-duration tests tend to show more variability in savings estimates due to smaller sample sizes. Longer experiments have larger sample sizes, so their results can generally be considered more reliable, all else being equal. \autoref{FIG9} shows the energy (top row) and cost (bottom) savings in residential (left column) and commercial (right) buildings, categorized by test durations of more or less than seven days. Shorter experiments not only show higher variability in savings estimates (as indicated by the taller boxes, which span the 25th to 75th percentiles), but also show systematically higher savings. For example, the duration-weighted average cost savings in residential buildings (lower left subplot) are 57\% for experiments shorter than one week and 22\% for longer experiments.

We conjecture that thermal mass effects could explain the systematically higher savings estimates in shorter experiments. In winter, for example, an advanced heating control system might let temperatures drift in unoccupied zones, leading to lower time-average air temperatures than observed under baseline control. Lower air temperatures would cause heat transfer from the building's thermal mass to the indoor air, effectively discharging the building's passive thermal energy storage. When the advanced control experiment ended, the baseline control method would return zone air temperatures to their original, higher values and heat would transfer from air back to thermal mass. In other words, the advanced control system would `inherit' fully charged thermal storage from the baseline control system and then leave the thermal storage partially depleted. Consequently, the baseline control system would later have to use more energy than usual to restore the thermal storage. Failure to account for these thermal mass effects would lead to more pronounced biases in the results of shorter experiments. Transient thermal mass effects typically last no more than a handful of days, so the bias would be small in an experiment that lasted a month or more. For an experiment that lasted only a few days, however, thermal mass artifacts could be significant, or even dominant.

To promote reliability of energy and cost savings estimates, all subsequent analyses in this paper use only savings estimates from experiments that lasted at least seven days. Shorter experiments are filtered out due to the risk that they overestimate savings. Filtering out short-duration tests removed 13 (16\%) of the 82 tests that remained after filtering out studies that used simulation-based benchmarks or controlled a small subset of building spaces.

\begin{figure}[!t]
    \centering
    \includegraphics[width=0.975\textwidth]{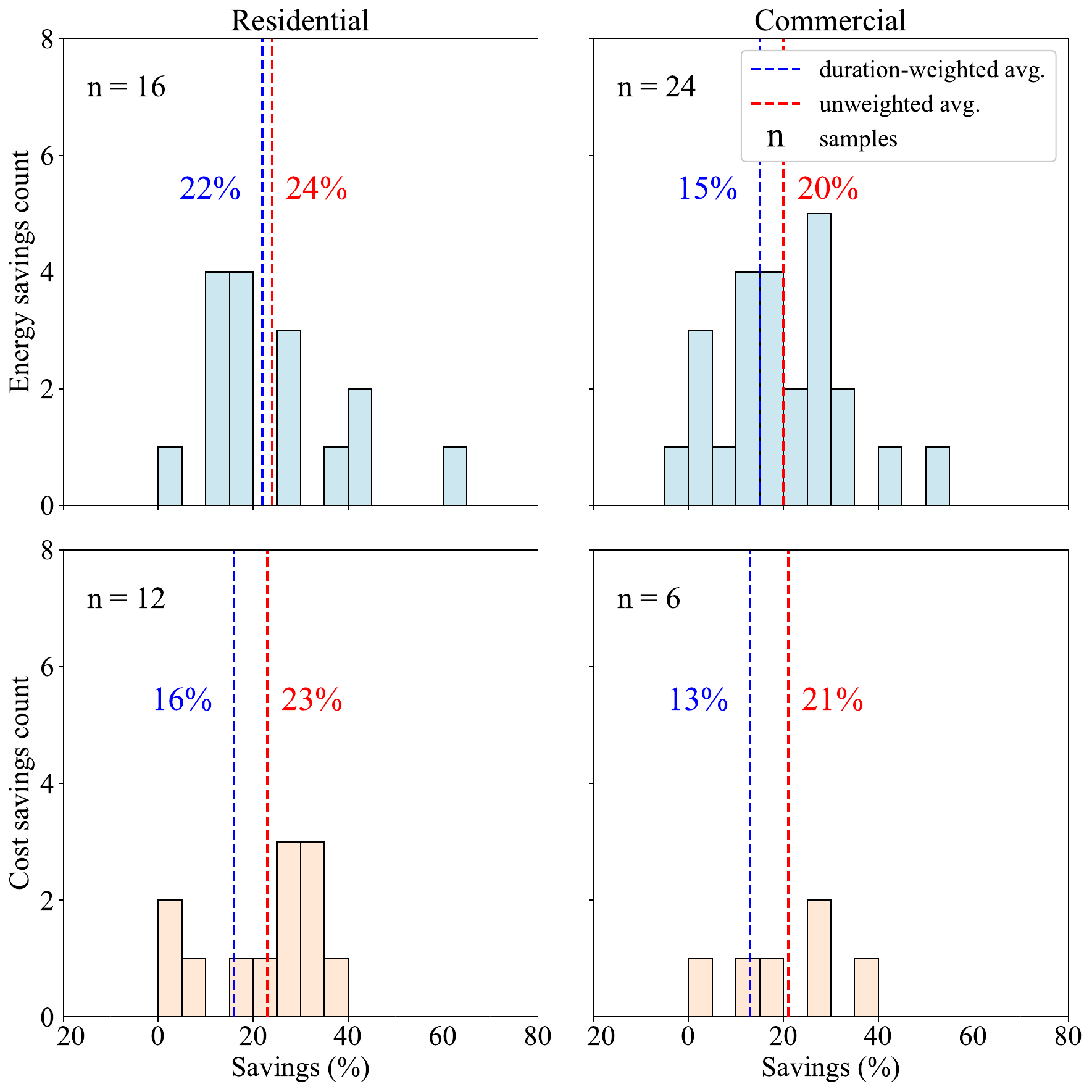}
    \caption{Histograms of energy (top) and cost (bottom) savings in residential (left) and commercial (right) tests after applying the experiment protocol filters.}
    \label{FIG10}
\end{figure}

\subsection{Energy and cost savings in filtered data}

Applying the three experiment protocol filters described in this section -- selecting for studies that used measurement-based benchmarking, controlled a whole building or a large subset of a building, and tested for at least seven days -- left 25 (48\%) and 28 (18\%) tests from the initial residential and commercial pools of 52 and 154 field tests, respectively. These tests are attributed to 9 out of 24 (37\%) residential papers and 21 out of 80 (26\%) commercial papers, or 29\% of the 104 total papers. We also excluded 16 of the 206 total tests due to incomplete reporting of experiment protocols. \autoref{FIG10} shows the energy and cost savings over these remaining 53 tests. The energy savings (top row of plots) ranged from -0.4\% to 60.4\%. The duration-weighted average energy savings were 22\% for residential buildings (left column) and 15\% for commercial buildings (right). The cost savings (bottom) ranged from 1.3\% to 39.0\%. The duration-weighted average cost savings were 16\% for residential buildings and 13\% for commercial buildings.

\section{Discussion}

\label{sec:discussion}

Here we discuss a number of potentially informative analyses we conducted that ended up being inconclusive due to small sample sizes or inconsistent reporting of relevant information (Section \ref{inconclusive}). We then summarize any available information on costs, labor, and other challenges related to deploying, operating, and maintaining MPC and RL systems (Section \ref{costs}). Based on the reviewed studies, we recommend experiment protocols and research directions for future field work (Section \ref{recommendations}). Lastly, we discuss limitations of this paper and opportunities for future literature reviews to address those limitations (Section \ref{limitations}).

\subsection{Inconclusive analyses}
\label{inconclusive}

In this section, we discuss our attempts to analyze how energy and cost savings from MPC and RL depend on the electricity rate structure, the building properties, the control architecture, and the climate zone and operational modes. Although studying these factors could help identify settings where MPC and RL are particularly effective, inconsistencies in reporting across studies made it difficult to draw robust conclusions.

\subsubsection{Electricity rate structure} 

We hypothesize that the complexity of the electricity rate structure substantially affects cost savings from MPC and RL. While human experts can design good heuristic control methods for simple economic contexts, such as time-invariant electricity prices, this task becomes more challenging under more complex incentives such as time-of-use pricing, critical peak pricing, day-ahead hourly pricing tied to wholesale markets, peak demand charges, or providing power grid reliability services such as spinning reserve or frequency regulation. Under more complex incentives, we therefore expect MPC and RL to show larger savings relative to baseline control. We attempted to test this hypothesis by categorizing studies into tiers of economic complexity and comparing cost savings (see Section \ref{ratestructure}), but after applying the experiment protocol filters discussed in Section \ref{sec:protocols}, the sample sizes were too small to draw statistically significant conclusions.

\subsubsection{Building properties} 

We hypothesize that building properties can substantially affect cost savings from MPC and RL. Thermal mass, for example, influences a building's capacity to store thermal energy, and as a result, a control system's ability to shift energy use across time. Similarly, insulation has a significant impact on HVAC performance by determining the extent of heat retention and thermal losses, which affect load-shifting potential. We attempted to test this hypothesis by categorizing buildings as low-to-moderate vs. moderate-to-high insulation levels and thermal masses (see Sections \ref{thermalmass} and \ref{insulation}). However, few papers reported this information, making accurate classification of studies very challenging. After applying the experimental filters outlined in Section \ref{sec:protocols}, sample sizes were too small to draw statistically significant conclusions.

\subsubsection{Control architecture} 

We investigated whether the control architecture -- for example, adjusting high-level setpoints vs. low-level actuator positions -- influences energy and cost savings from MPC and RL (see Section \ref{control}). We attempted to test hypothesis that high- and low-level control architectures lead to different savings distributions. However, after applying the experiment protocol filters defined in Section \ref{sec:protocols}, the sample sizes were too small to draw statistically significant conclusions.

\subsubsection{Climate and operating mode} 

We investigated the influence of climate conditions and HVAC operating modes (heating, cooling, or both) on energy and cost savings from advanced control. Specifically, we categorized studies based on climate zones, heating and cooling degree-days, diurnal temperature variations, and operating modes (see Sections \ref{weather} and \ref{operationalmode}). However, after applying the experiment protocol filters defined in Section \ref{sec:protocols}, no clear trends emerged. This is possibly due to inconsistencies in how studies reported climate data and operational modes, to variations in building properties and HVAC equipment across studies, or to the uneven geographic distribution of field studies.

\subsection{Costs, labor, and other challenges}
\label{costs}

After we applied the experiment protocol filters defined in Section \ref{sec:protocols}, the duration-weighted average cost savings were 16\% in residential buildings and 13\% in commercial buildings. While these cost savings are potentially attractive, making the economic case for advanced HVAC control also requires understanding the costs of deployment, operation, and maintenance. Only 13 papers provide information on these costs, which we organize into four categories: instrumentation, data and communication, labor, and cost-benefit analysis. This section collects what little cost information has been reported.

\subsubsection{Instrumentation}

Instrumentation costs include expenses associated with devices for sensing, actuation, communication, and computing. With a focus on residential buildings, Finck et al. \cite{Finck2020-ea} reported hardware costs including a data logger (\$2,800, about \$3,450 in 2025), a flow meter (\$200+, about \$246+ in 2025), and a smart power meter (\$200+, about \$246+ in 2025). Wang et al. \cite{Wang2023-ss} described a low-cost plug-and-play system priced at \$50 (about \$52.5 in 2025) that used two smart wall sockets and two temperature sensors. Pergantis et al. \cite{Pergantis2025} detailed an approach that used a power meter (\$100) and smart thermostats with Wi-Fi control via manufacturer application programming interfaces. With a focus on commercial buildings, Goyal et al. \cite{Goyal2015-ce} reported a \$215 (about \$280 in 2025) occupancy sensor for an occupancy-based setback controller. Sturzenegger et al. \cite{Sturzenegger2016-vd} documented additional instrumentation costs, such as new room temperature sensors, solar radiation sensors, and blinds actuators, costing 3,000 to 6,000 CHF (about \$3,400 to \$6,800 in 2025). Ham et al. \cite{Ham2024-vg} noted that their MPC infrastructure cost only \$975 (about \$1,200 in 2025) in hardware. They additionally spent \$7,330 (about \$9,020 in 2025) on instrumentation for measurement and verification, a cost mostly for research purposes rather than routine operation.

\subsubsection{Data and networking}

Data and networking costs include expenses for off-site data sources, such as weather forecasts, and on-site information technology services required for reliable data access and communication. Pergantis et al. \cite{Pergantis2025} used manufacturer-provided application programming interfaces, which cost nothing for their research purposes but might entail subscription fees for deployment at scale. Sturzenegger et al. \cite{Sturzenegger2016-vd} reported an annual subscription to weather forecast data that ranged from 100 to 600 CHF (about \$115 to \$680 in 2025).

\subsubsection{Labor}

Labor costs include the human effort required to deploy, operate, and maintain an advanced HVAC control system. In residential buildings, Pergantis et al. \cite{Pergantis2024a} reported that MPC deployment required 190 engineer-days in total, delineated between 150 days for one-time tasks (which would not have to be repeated for each new deployment) and 40 days for recurring tasks (which would have to be repeated for each new deployment) such as device installation, calibration, and troubleshooting. Ongoing maintenance activities are also labor intensive, as noted by Knudsen et al. \cite{Knudsen2021-qw}, who stressed the need for periodic model recalibration, and Svetozarevic et al. \cite{Svetozarevic2022-zk}, who emphasized the need for continuous commissioning. In commercial buildings, Sturzenegger et al. \cite{Sturzenegger2016-vd} highlighted high labor demands for model development and for integrating and commissioning devices. Ruano et al. \cite{Ruano2016-uf} described a one-week installation process for hardware setup and system integration. Granderson et al. \cite{Granderson2018-gp} reported that installation, information technology approvals, troubleshooting, and stability monitoring can require up to three days of staff time. Blum et al. \cite{Blum2022-dc} quantified these efforts in a commercial MPC deployment. They estimated 239 person-days total, with 70 person-days for preparation, 79 for model development and integration, 30 for controller software development, and 60 for deployment (commissioning, building management system integration, and maintenance).

\subsubsection{Cost-benefit analysis}

Project finance metrics -- such as simple payback, benefit-to-cost ratio, net present value, internal rate of return, etc. -- provide key information to business managers who decide whether to deploy advanced HVAC control systems. In residential buildings, Thorsteinsson et al. \cite{Thorsteinsson2023-ct} estimated a cost recovery of €1,200 (about \$1,320 in 2025) over 10 years. In commercial buildings, Ham et al. \cite{Ham2024-vg} reported a payback period of 13 heating months for MPC infrastructure. Sturzenegger et al. \cite{Sturzenegger2016-vd} cautioned that high deployment and engineering costs may outweigh potential savings unless improvements in automation and modeling frameworks are achieved.

\subsection{Recommendations for future field studies}
\label{recommendations}

\subsubsection{Improve experiment protocols} 

We recommend that future field demonstrations 1) compare advanced control performance to the measured performance of a baseline controller in the same building or a nearby, similar building; 2) control a whole building, control a large subset of a building that is well-insulated from adjacent zones, or account in savings calculations for heat transfer between controlled zones and adjacent zones; and 3) test advanced control for at least seven days, and ideally for one month or longer. As discussed in Section \ref{sec:protocols}, past field demonstration results suggest that deviating from any of these three protocols can lead to overestimating savings from advanced control. Researchers can encourage the use of these protocols by implementing them in their own field demonstrations and by checking for them in editorial and peer-review processes for field demonstration submissions to research journals and conferences.

\subsubsection{Focus on costs} 

Making the economic case for advanced HVAC control requires understanding not only its benefits, but also its costs. While all 104 of the reviewed field studies reported benefits, such as energy savings or utility bill savings, only 13 reported any information related to the costs of deploying, operating, and maintaining an advanced control system. We recommend that future field studies report as much information as possible about the costs, labor, and other challenges associated with deployment, operations, and maintenance. We also recommend that future field studies distinguish between one-time costs (such as training junior researchers to implement advanced control, or selecting appropriate instrumentation and algorithms) vs. recurring costs (such as instrumenting a building, establishing data access and communication, on-boarding a building into a pre-existing software system, commissioning systems, and tuning controllers) that a company would need to repeat for each new building. In our view, given the current state of the advanced HVAC control field, the research community should refocus its efforts almost exclusively on reducing the recurring costs described above. Breakthroughs could include algorithms that need radically less sensing and communication infrastructure, or algorithms and infrastructure that deploy and commission far faster than existing approaches and need far less ongoing attention to operate and maintain.

\subsubsection{Report the context} 

As discussed in Section \ref{inconclusive}, advanced HVAC control may prove more effective in some contexts than in others. For example, we conjecture that advanced control should be particularly effective under complex electricity rate structures and in thermally massive buildings. Some control architectures (such as adjusting high-level setpoints vs. low-level actuator positions) may prove more effective than others. Climate considerations, such as diurnal variations in temperature and humidity, also might matter. Unfortunately, we found that reporting of these contextual data was too sparse and inconsistent to enable drawing rigorous conclusions. Therefore, we recommend that future field studies clearly report at least the following contextual data:
\begin{itemize}
    \item Whether the electricity rate structure includes a time-invariant price, time-of-use prices with two or more price tiers, hourly prices, critical peak prices, peak demand charges, or incentives for distribution-level demand response or transmission-level ancillary services.

    \item Estimates of the date of construction or latest renovation, the thermal resistance of the building envelope, the outdoor air infiltration rate, and the thermal capacitance of the building materials. If these data are unavailable, we recommend that researchers provide qualitative assessments of whether the building is old or new; is insulated poorly or well; is drafty or tight; and is thermally light or heavy.
    
    \item Whether the researchers adjusted high-level setpoints or low-level actuator positions.

    \item Characterization of the climate zone (e.g., hot, mixed, or cold; dry, humid, or marine).
    
\end{itemize}

To facilitate reporting of the contextual variables described above, we provide a checklist of experiment protocols and contextual data in the form of an open-access online spreadsheet \cite{field-demonstrations-HVAC-MPC-RL}. We recommend that a) researchers who conduct future field demonstrations populate a copy of this spreadsheet and attach it as a digital artifact to their journal or conference submissions; and b) editors and peer-reviewers who interact with field demonstration submissions to journals or conferences check that the authors have attached this spreadsheet or provided comparable information in another format.

\subsubsection{Expand the hardware scope} 

Almost all of the reviewed field studies controlled HVAC systems only, ignoring interactions with other equipment. We recommend that future field demonstrations coordinate HVAC systems alongside other distributed energy resources (DERs), such as electric vehicles (EVs), stationary batteries, thermal storage, or solar photovoltaics (PV). We conjecture that jointly coordinating a variety of DERs under one control system, rather than separately controlling each class of equipment, will substantially improve the overall business case for advanced control.

\subsubsection{Explore emerging objectives} 

While many simulation studies have explored the potential of HVAC equipment to actively participate in power grid operations, few field studies have evaluated these opportunities in real, grid-interactive buildings. We recommend that future field studies control HVAC equipment (possibly alongside other DERs) to balance a range of objectives, including both traditional goals such as improving energy efficiency and emerging goals such as peak shaving, arbitraging day-ahead or real-time wholesale energy prices, and providing power grid reliability services such as spinning reserve, frequency regulation, and voltage support.

\subsection{Limitations and future literature reviews}
\label{limitations}

While this paper reviewed all peer-reviewed field demonstrations of MPC and RL for residential and commercial HVAC of which the authors are aware, it has several limitations. To keep the analysis tractable, we limited the scope of our quantitative analyses to peer-reviewed journal articles and refereed conference papers, excluding white papers, theses, dissertations, and technical reports. Future literature review papers could expand our open-access dataset \cite{field-demonstrations-HVAC-MPC-RL} to include these sources.

This paper also focused primarily on MPC and RL, two popular algorithms for advanced HVAC control. Future work could expand the scope to include field demonstrations of other advanced control methods, comparing their benefits and costs to those of MPC and RL. Some of these methods include open-loop optimal control \cite{morris1994experimental, Clarke2002-go, Lee2006-lj, lee2008model}, fuzzy logic control \cite{kang2015fuzzy, berouine2019fuzzy}, rule-based control \cite{lee2008data, may2013experimental, Brooks2015-zz, bursill2019experimental, bursill2020multi, Kong2022-mj}, robust control, adaptive control, and hybrid approaches that combine optimization and learning-based methods. Furthermore, given the growing interest in developing model-free control methods, variants of MPC such as DeePC \cite{Lian2023-jx, Yin2024-xh, Shi2025-jt} could be investigated further.

The analysis in this paper was also limited by sparse and inconsistent reporting on electricity rate structures, building properties, control architectures, climate data, and other contextual information. If future field studies standardize reporting of these data, future literature review papers could revisit the hypotheses discussed in Section \ref{inconclusive} to identify contexts in which advanced HVAC control is particularly effective.

\section{Conclusion}
\label{sec:conclusion}

This paper analyzed 104 peer-reviewed field demonstrations of MPC and RL for residential and commercial HVAC. We found that 71\% of reviewed papers used experiment protocols that may result in unreliable savings estimates. Over the remaining 29\%, the duration-weighted average utility bill savings were 16\% in residential buildings and 13\% in commercial buildings. While these savings are potentially attractive, few studies reported on costs, labor, or other challenges related to deploying, operating, and maintaining advanced control systems. To help make the economic case for advanced HVAC control, and thereby accelerate real-world adoption, we recommend that the research community pivot its efforts toward characterizing and reducing the costs of deployment, operation, and maintenance. We also recommend controlling HVAC systems for emerging objectives, possibly alongside other distributed energy resources, to improve the overall economic case for advanced control.

\section*{CRediT authorship contribution statement}

{\bf Arash J. Khabbazi:} Conceptualization, Methodology, Software, Validation, Formal Analysis, Investigation, Data Curation, Writing - Original Draft, Writing - Review \& Editing, Visualization, Supervision. {\bf Elias N. Pergantis:} Methodology, Formal Analysis, Investigation, Writing - Original Draft, Writing - Review \& Editing. {\bf Levi D. Reyes Premer:} Investigation, Writing - Review \& Editing. {\bf Panagiotis Papageorgiou:} Investigation, Writing - Review \& Editing. {\bf Alex H. Lee:} Investigation, Writing - Review \& Editing. {\bf James E. Braun:} Writing - Review \& Editing. {\bf Gregor P. Henze:} Conceptualization, Writing - Review \& Editing. {\bf Kevin J. Kircher:} Conceptualization, Methodology, Validation, Writing - Review \& Editing, Supervision, Project Administration, Funding Acquisition.

\section*{Declaration of competing interest}

The authors declare that they have no known competing financial interests or personal relationships that could have appeared to influence the work reported in this paper.

\section*{Data availability}

Data are publicly available in \cite{field-demonstrations-HVAC-MPC-RL}.

\section*{Acknowledgments}
\label{sec:acknowledgments}

We sincerely thank the following researchers for responding to our requests for further information about their field demonstrations: José Domingo Álvarez Hervás (University of Almería), David Blum (Lawrence Berkeley National Laboratory [LBNL]), Hanmin Cai (Swiss Federal Laboratories for Materials Science and Technology), Roel De Coninck (dnergy), Ján Drgoňa (Johns Hopkins University), Svenne Freund (ROM Technik), Jessica Granderson (LBNL), Sang woo Ham (LBNL), Lieve Helsen (KU Leuven), Hao Huang (Frasers Property Limited), Yilin Jiang (Pacific Northwest National Laboratory), Donghun Kim (LBNL), Michael Dahl Knudsen (Aarhus University), David Lindelöf (Google), Iakovos Michailidis (The Centre for Research \& Technology Hellas), Avisek Naug (Hewlett Packard Labs), Zheng O'Neill (Texas A\&M University), Marco Pritoni (LBNL), Etienne Saloux (CanmetENERGY), Jicheng Shi (École Polytechnique Fédérale de Lausanne), Jan Široký (Energy Twin), Phillip Stoffel (E.ON Energy Research Center - RWTH Aachen University), Jacopo Vivian (University of Padua), Xuezheng Wang (Syracuse University), Sicheng Zhan (Massachusetts Institute of Technology), and Kun Zhang (École de Technologie Supérieure Montréal). Thanks as well to Ettore Zanetti (LBNL) for helpful discussion. We gratefully acknowledge scholarship support from The American Society of Heating, Refrigerating and Air-Conditioning Engineers Grant-In-Aid Award Program (A.J. Khabbazi; E.N. Pergantis; L.D. Reyes Premer), scholarship support from The Alexander S. Onassis Foundation (E.N. Pergantis; P. Papageorgiou), and Graduate Research Fellowships Program support from the U.S. National Science Foundation (L.D. Reyes Premer).

\appendix
\section{Nomenclature}
\label{nomenclature}

\begin{tabular}{l | l}
AHU & air handling unit \\
DeePC & data-enabled predictive control \\
DER & distributed energy resource \\
HP & heat pump \\
HVAC & heating, ventilation, and air-conditioning \\
MPC & model predictive control \\
OCC & occupant-centric control \\
PV & photovoltaics \\
RL & reinforcement learning \\
TABS & thermally-activate building system \\
VAV & variable air volume \\
VRF & variable refrigerant flow \\
\end{tabular}

\section{Field demonstration summary tables}
\label{sec:appendixa}
\clearpage % Ensure a fresh page

\onecolumn % Switch to single-column mode
\begin{landscape} % Switch to landscape mode
\fontsize{8pt}{10pt}\selectfont
\begin{longtable}{@{}p{0.15cm} p{0.5cm} p{0.75cm} 
>{\raggedright\arraybackslash}p{1.75cm} 
>{\raggedright\arraybackslash}p{3.5cm} 
>{\raggedright\arraybackslash}p{4.5cm} 
p{1cm} 
>{\raggedright\arraybackslash}p{2.5cm} 
>{\raggedright\arraybackslash}p{5cm}@{}}
\caption{Overview of field demonstrations of MPC and RL for HVAC in residential buildings.}
\label{tab:res} \\
\toprule
\textbf{\#} & \textbf{Study (Year)} & \textbf{Control Method} & \textbf{Location} & \textbf{Building Description} & \textbf{HVAC System} & \textbf{Test Days} & \textbf{Objective(s)} & \textbf{Reported Outcome(s) \newline vs. Benchmarking Method} \\
\midrule
\endfirsthead

\multicolumn{9}{c}%
{\tablename\ \thetable\ -- \textit{Continued from previous page}} \\
\toprule
\textbf{\#} & \textbf{Study (Year)} & \textbf{Control Method} & \textbf{Location} & \textbf{Building Description} & \textbf{HVAC System} & \textbf{Test Days} & \textbf{Objective(s)} & \textbf{Reported Outcome(s) \newline vs. Benchmarking Method} \\
\midrule
\endhead

\bottomrule
\multicolumn{9}{r}{\textit{Continued on next page}} \\
\endfoot

\bottomrule
\endlastfoot

% done - checked
1 & \cite{Pedersen2013-ho} \newline 2013 & MPC & Denmark & four single-family detached houses & radiant floor heating with ground-to-water heat pumps (HPs) & 152 & minimize cost & 9.2\% cost savings vs. measured baseline \\

% done - checked
2 & \cite{Dong2014-lv} \newline 2014 & MPC & Pittsburgh, PA, \newline USA & a sustainable house, visitor location & radiant floor heating and forced-air cooling with fixed settings & 10 & minimize energy & 17.8--30.1\% energy savings vs. simulated baseline \\

% done - checked
3 & \cite{Lindelof2015-wq} \newline 2015 & MPC & Switzerland and Germany & eight single-family houses and two apartments & radiant with water distribution; sources include wood pellets, HPs, gas, and oil & 27--138 & minimize energy & 3.9--60.4\% energy savings vs. measured baseline \\

% done - checked
4 & \cite{Leurs2016-mq} \newline 2016 & RL & Leuven, Belgium & living lab setup featuring a test room & forced-air integrated with PV system & 3 & maximize solar self-utilization & reduced PV peak power injection and synchronized cooling with PV generation vs. measured baseline \\

% done - checked
5 & \cite{Zong2017-se} \newline 2017 & MPC & Roskilde, Denmark & single family detached house & radiant water-based heating with electro-valve radiators & 10 & minimize energy and cost; shift loads to off-peak periods & -3.2--5\% energy savings, 3--11.3\% cost savings, and effective load shift during off-peak hours vs. simulated baseline \\

% done - checked
6 & \cite{Afram2017-gy} \newline 2017 & MPC & Vaughan, ON, Canada & four-story sustainable house, including a basement, with an adjacent in-law suite & hybrid system with radiant heating, forced-air cooling, and a ground source HP with a buffer tank & 22 & minimize energy and cost & -4.1--52.2\% energy savings and 8.5--56.1\% cost savings vs. simulated baseline \\

% done - checked
7 & \cite{Bunning2020-rp} \newline 2020 & MPC & Dübendorf, Switzerland & residential module in a sustainable demonstrator building & air-to-water HP with hydronic radiant ceilings and room-specific valves & 6 & minimize energy & 24.9\% energy savings vs. measured baseline \\

% done - checked
8 & \cite{Finck2020-ea} \newline 2020 & MPC & near Amstelveen, Netherlands & detached four-story low-energy house & hydronic floor heating supplied by an HP powered by PV thermal panels & 1 & minimize cost & 12--15\% cost savings vs. measured baseline \\

% done - checked
9 & \cite{Kurte2020-my} \newline 2020 & RL & Knoxville, TN, USA & detached energy-efficient house & two-zone air-to-air conditioning with two-stage compressor and variable-speed fan & 5 & minimize cost & 11--21\% cost savings vs. simulated baseline \\

% done - checked
10 & \cite{Knudsen2021-qw} \newline 2021 & MPC & Trondheim, Norway & test house at a university & hydronic floor heating from a resistive-heated tank, with mechanical ventilation & 14 & minimize cost & 22.5\% cost savings vs. simulated baseline \\

% done - checked
11 & \cite{Svetozarevic2022-zk} \newline 2022 & RL & Dübendorf, Switzerland & residential module in a sustainable demonstrator building & radiant heating with HP and emulated electric vehicle integration & 5 & minimize energy & 27\% energy savings vs. measured baseline \\

% done - checked
12 & \cite{Bunning2022-fz} \newline 2022 & MPC & Dübendorf, Switzerland & residential module in a sustainable demonstrator building & hydronic loop with HPs, variable-speed pump, room valves, and ceiling radiant systems & 35--41 & minimize energy & 26--49\% energy savings vs. measured baseline \\

% done - checked
13 & \cite{Vivian2022-wm} \newline 2022 & MPC & Piacenza, Italy & laboratory building & air-to-water HP with fan coils, air extractors, and simulated PV panels & 5 & minimize cost & 10--17\% cost savings vs. measured baseline \\

% done - checked
14 & \cite{Brown2023-ea} \newline 2023 & MPC & Ottawa, ON, Canada & two-story research facility within a university & hydronic floor system for heating and cooling & 13 & minimize energy & 40.6\% energy savings vs. measured baseline \\

% done - checked
15 & \cite{Wang2023-ss} \newline 2023 & MPC & Shenzhen, China & two-room apartment with living room & forced-air with fixed-speed, single-stage mini-split air conditioners & 12 & minimize cost & 22.1--26.8\% cost savings vs. measured baseline \\

% done - checked
16 & \cite{Vallianos2024-ru} \newline 2023 & MPC & Shawinigan, QC, Canada & two-story detached house with a basement & radiant heating with electric baseboards and smart thermostats & 3 & minimize cost and shift load & 55\% cost savings and 71\% high-price energy reduction vs. measured baseline \\

% done - checked
17 & \cite{Thorsteinsson2023-ct} \newline 2023 & MPC & Zealand, Denmark & detached low-energy house & hydronic variable-speed air-to-water HP for space heating and domestic hot water & 10--50 & minimize cost & 2.3--32.8\% cost savings vs. measured baseline \\

% done - checked
18 & \cite{Brown2023-sn} \newline 2023 & MPC & Ottawa, ON, Canada & two-story research facility within a university & hydronic floor system for heating and cooling & 182 & minimize energy & runtime optimized; savings inferred vs. measured baseline \\

% done - checked
19 & \cite{Yin2024-xh} \newline 2024 & MPC \newline (DeePC) & Dübendorf, Switzerland & residential module in a sustainable demonstrator building & hydronic loop with HPs, variable-speed pump, room valves, and ceiling radiant systems & 5 & minimize energy & experimental savings unreported; simulated energy savings 12--27\% vs. simulated baseline \\

% done - checked
20 & \cite{Jiang2024-xj} \newline 2024 & MPC & Norman, OK, and Miami, FL, USA & nine single-family houses & central forced-air systems with ACs designed for cooling & 2--24 & minimize energy and cost & -29.8--91.1\% energy savings and -13.5--87.1\% cost savings vs. measured baseline \\

% done - checked
21 & \cite{Pergantis2024a} \newline 2024 & MPC & West Lafayette, IN, USA & detached full-electric single-family home & forced-air variable-speed air-to-air HP with staged electric resistance backup heating & 33 & minimize energy and cost & 19\% energy savings and 28\% cost savings vs. measured baseline \\

% done - checked
22 & \cite{Shamachurn2024-pi} \newline 2024 & MPC & Reduit, Mauritius, Africa & single-story, two-zone concrete building for tropical climate testing & forced-air with a portable AC, electric fan, and air extractors, with adjustable speed settings & $<1$ & minimize temperature tracking error & improved temperature tracking and efficiency vs. measured baseline \\

% done - checked
23 & \cite{Pergantis2024b} \newline 2024 & MPC & West Lafayette, IN, USA & detached full-electric single-family home & forced-air variable-speed air-to-air HP with staged electric resistance backup heating & 38 & minimize cost & -5--32.5\% cost savings vs. measured baseline \\

% done - checked
24 & \cite{Pergantis2025} \newline 2025 & MPC & West Lafayette, IN, USA & detached full-electric single-family home & forced-air variable-speed air-to-air HP with staged electric resistance backup heating & 31 & whole home hard current draw constraint & maintained peak current draw below 100 A (no benchmark) \\

\end{longtable}
\end{landscape}
%\twocolumn % Switch back to two-column mode
\clearpage

\clearpage % Ensure a fresh page
\onecolumn % Switch to single-column mode
\begin{landscape} % Switch to landscape mode
\fontsize{8pt}{10pt}\selectfont
\begin{longtable}{@{}p{0.15cm} p{0.5cm} p{0.75cm} 
>{\raggedright\arraybackslash}p{1.75cm} 
>{\raggedright\arraybackslash}p{3.5cm} 
>{\raggedright\arraybackslash}p{4.5cm} 
p{1cm} 
>{\raggedright\arraybackslash}p{2.5cm} 
>{\raggedright\arraybackslash}p{5cm}@{}}
\caption{Overview of field demonstrations of MPC and RL for HVAC in commercial buildings.}
\label{tab:com} \\
\toprule
\textbf{\#} & \textbf{Study (Year)} & \textbf{Control Method} & \textbf{Location} & \textbf{Building Description} & \textbf{HVAC System} & \textbf{Test Days} & \textbf{Objective(s)} & \textbf{Reported Outcome(s) \newline vs. Benchmarking Method} \\
\midrule
\endfirsthead

\multicolumn{9}{c}%
{\tablename\ \thetable\ -- \textit{Continued from previous page}} \\
\toprule
\textbf{\#} & \textbf{Study (Year)} & \textbf{Control Method} & \textbf{Location} & \textbf{Building Description} & \textbf{HVAC System} & \textbf{Test Days} & \textbf{Objective(s)} & \textbf{Reported Outcome(s) \newline vs. Benchmarking Method} \\
\midrule
\endhead

\bottomrule
\multicolumn{9}{r}{\textit{Continued on next page}} \\
\endfoot

\bottomrule
\endlastfoot

% done - checked
1 & \cite{Henze2005-ed} \newline 2005 & MPC & Ankeny, IA, US & eight office rooms at a test facility & forced-air fixed-speed with air-cooled chillers, ice storage, and air handling units (AHUs) & 4 & minimize cost & experimental savings negligible; simulated cost savings 17--27\% vs. simulated baseline \\

% done - checked
2 & \cite{Liu2006-ik} \newline 2006 & RL & Ankeny, IA, US & eight office rooms at a test facility & forced-air fixed-speed with air-cooled chillers, ice storage, and AHUs & 6 & minimize cost & 8.3\% costing savings vs. simulated baseline \\

% done - checked
3 & \cite{Kolokotsa2009-ew} \newline 2009 & MPC & Chania, Crete, Greece & laboratory in a university building & forced-air heating and cooling via split-type air conditioning & 3 & minimize energy & experimental savings unreported vs. measured baseline \\

% done - checked
4 & \cite{Liao2010-sa} \newline 2010 & MPC & Garston, England & office building with office, meeting rooms, and lecture theaters & radiant heating with thermostatic valves & 45 & minimize energy & experimental savings unreported; simulated energy savings 9--32.1\% vs. simulated baseline \\

% done - checked
5 & \cite{Privara2011-xl} \newline 2011 & MPC & Prague, Czech Republic & university building with seven control blocks & hydronic ceiling radiant heating with embedded beams, heat exchanger, and valve control & 45 & minimize energy and cost & 20.5--29\% energy savings and 29\% cost savings vs. measured baseline \\

% done - checked
6 & \cite{Siroky2011-tb} \newline 2011 & MPC & Prague, Czech Republic & university building with seven control blocks & hydronic ceiling radiant heating with embedded beams, heat exchanger, and valve control & 7--49 & minimize energy & 16.24--28.74\% energy savings vs. measured baseline \\

% done - checked
7 & \cite{Castilla2011-bz} \newline 2011 & MPC & Almería, Spain & office space in a bioclimatic research facility & forced-air with variable fan speed and chilled water coils & NR\footnote{Not reported} & minimize energy & experimental savings unreported vs. simulated baseline \\

% done - checked
8 & \cite{Aswani2012-hu} \newline 2012 & MPC & Berkeley, CA, US & ground-floor university building & forced-air, single-stage HP with single-speed operation & 3 & minimize energy & 28--66\% energy savings vs. simulated baseline \\

% done - checked
9 & \cite{Ma2012-nc} \newline 2012 & MPC & Merced, CA, US & university building & forced-air centralized system with chillers, thermal energy storage, and distribution loops & 5 & minimize cost and maximize coefficient of performance & \$1,280 daily cost savings and 19.1\% coefficient of performance improvement vs. measured baseline \\

% done - checked
10 & \cite{Gayeski2012-tm} \newline 2012 & MPC & Atlanta, GA and Phoenix, AZ, US & test chamber simulating office conditions with another climate chamber & thermally-activate building system (TABS) with radiant cooling by low-lift chilled water & 7 & minimize energy & 19--25\% energy savings vs. measured baseline \\

% done - checked
11 & \cite{Ferreira2012-cr} \newline 2012 & MPC & Algarve, Portugal & four rooms in a university building & forced-air variable refrigerant flow (VRF) systems with ducted indoor and outdoor units per zone & 1--2 & minimize energy & 41--77\% energy savings vs. measured baseline \\

% done - checked
12 & \cite{Bengea2014-ia} \newline 2014 & MPC & Champaign, IL, US & office building & forced-air with variable air volume (VAV) dual-duct multi-zone AHU & 5 & minimize energy & 60--85\% energy savings vs. measured baseline \\

% done - checked
13 & \cite{West2014-vt} \newline 2014 & MPC & Newcastle and Melbourne, Australia & three-story office buildings & forced-air with AHUs, chillers, combined heat and power, and boilers & 5--25 & minimize energy and cost & 14--32\% energy savings and 16.9\% cost savings vs. measured baseline \\

% done - checked
14 & \cite{Castilla2014-xb} \newline 2014 & MPC & Almería, Spain & office space in a bioclimatic research facility & forced-air with variable fan speed and chilled water coils & 15 & minimize energy & 53\% energy savings vs. measured baseline \\

% done - checked
15 & \cite{Vana2014-cu} \newline 2014 & MPC & Hasselt, Belgium & three-floor office building, underground garage and under-roof apartment excluded & hydronic ground-coupled HPs with TABS, AHUs, floor heating, and seasonal thermal storage & 25 & minimize energy & 11.4--22.8\% energy savings vs. measured baseline \\

% done - checked
16 & \cite{Maasoumy2014-jn} \newline 2014 & MPC & Berkeley, CA, US & university building & forced-air variable-speed fans and variable frequency drives & $<1$ & maximize flexibility & 25\% fan power consumption variation and up to 20\% downward flexibility vs. measured baseline \\

% done - checked
17 & \cite{Preglej2014-pk} \newline 2014 & MPC & Graz, Austria & research facility simulating real-world conditions & forced-air with fans, coils, humidifiers, dampers, and filters & $<1$ & minimize energy & 44\% energy savings vs. measured baseline \\

% done - checked
18 & \cite{Parisio2014-nx} \newline 2014 & MPC & Stockholm, Sweden & ground-floor university laboratory & forced-air ventilation, district radiator heating, and inlet coil cooling & 2 & minimize energy & experimental savings unreported vs. measured baseline \\

% done - checked
19 & \cite{Parisio2014-xg} \newline 2015 & MPC & Stockholm, Sweden & ground-floor university laboratory & radiant heating via district-heated water radiators & 1 & minimize energy & 31--33\% energy savings vs. measured baseline \\

% done - checked
20 & \cite{Goyal2015-ce} \newline 2015 & MPC & Gainesville, FL, US & office room in a university building & forced-air via AHU, with VAV terminal and reheat for zone control & 1 & minimize energy & 40\% energy savings vs. measured baseline \\

% done - checked
21 & \cite{Zeng2015-eb} \newline 2015 & MPC & Ankeny, IA, US & four office rooms at a test facility & forced-air with one AHU connected to multiple VAVs & 1 & minimize energy & 2.2--17.2\% energy savings vs. simulated baseline \\

% done - checked
22 & \cite{Huang2015-fc} \newline 2015 & MPC & Adelaide, Australia & one zone in the check-in hall of an airport & forced-air cooling with chillers, AHUs, and variable-speed pumps & 1 & minimize cost & 13\% cost savings vs. measured baseline \\

% done - checked
23 & \cite{Li2015-mm} \newline 2015 & MPC & Philadelphia, PA, US & one-third of a medium-sized four-story office building & forced-air with three AHUs, VAVs with direct expansion cooling and gas boiler heating & 20 & minimize energy & 33\% energy savings vs. measured baseline \\

% done - checked
24 & \cite{Kim2015-ht} \newline 2015 & MPC & Knoxville, TN, US & basketball court within a gym & forced-air with four two-stage rooftop units (RTUs) & 4 & minimize energy and peak demand & 8\% energy savings and 43\% peak demand reduction vs. measured baseline \\

% done - checked
25 & \cite{Sturzenegger2016-vd} \newline 2016 & MPC & Allschwil, Switzerland & office building with five office floors and ground-floor kitchen and restaurant & hybrid system with AHU, TABS, gas boiler, dry cooling tower, and corner radiators & 10--98 & minimize cost & experimental savings unreported; simulated cost savings 17\% vs. simulated baseline \\

% done - checked
26 & \cite{De_Coninck2016-gt} \newline 2016 & MPC & Brussels, Belgium & two-story office building & hybrid system with air-source HPs, boiler, and hydronic loop & 15 & minimize energy and cost & 20--30\% energy savings and 34--40\% cost savings vs. measured baseline \\

% done - checked
27 & \cite{Ruano2016-uf} \newline 2016 & MPC & Faro, Portugal & three rooms in a university building & forced-air VRF system with air-cooled outdoor unit and ceiling-concealed indoor units & $<1$ & minimize energy & 50--51\% energy savings vs. measured baseline \\

% done - checked
28 & \cite{Costanzo2016-vp} \newline 2016 & RL & Genk, Belgium & laboratory space for energy testing & ductless air conditioner units with basic monitoring & 16 & minimize cost & experimental savings unreported (no benchmark); simulation performance within 90\% of mathematical optimum vs. simulated baseline \\

% done - checked
29 & \cite{Fabietti2018-qy} \newline 2016 & MPC & Lausanne, Switzerland & four office rooms in a university research lab & radiant heating with commercial electric heater with hot water radiators disabled & $<1$ & frequency regulation evaluation & improved frequency regulation vs. simulated baseline \\

% done - checked
30 & \cite{Vrettos2018-hy} \newline 2016 & MPC & Berkeley, CA, US & thermally identical test cells in a research facility & forced-air with variable speed and dedicated AHUs & 1 & frequency regulation evaluation & improved frequency regulation vs. measured baseline \\

% done - checked
31 & \cite{Gorecki2017-af} \newline 2017 & MPC & Lausanne, Switzerland & five offices in an office building & electric heaters with direct modulation & 1 & frequency regulation evaluation & improved frequency regulation vs. simulated baseline \\

% done - checked
32 & \cite{Schmidt2017-xe} \newline 2017 & MPC, \newline RL & Granada, Spain & three-story school & hydronic heating with biomass boiler serving multiple circuits & 43 & minimize energy and improve comfort & 33\% energy savings and comfort improvement vs. measured baseline \\

% done - checked
33 & \cite{Stauffer2017-al} \newline 2017 & MPC & Neuchâtel and Winterthur, Switzerland & conference room and office building & forced-air with AHUs, radiators, and rotating heat exchanger & 10--38 & minimize cost & 21--23\% cost savings vs. measured baseline \\

% done - checked
34 & \cite{Hilliard2017-az} \newline 2017 & MPC & Halifax, NS, Canada & five-story university building & forced-air with VAV reheat, HPs, district heating, and rooftop cooling tower & 113 & minimize energy & 29--63\% energy savings vs. measured baseline \\

% done - checked
35 & \cite{Killian2018-nv} \newline 2018 & MPC & Salzburg, Austria & two floors of a five-story university office building & hybrid system with TABS (radiant) and fan coils (forced-air) & 49 & minimize energy & experimental savings unreported; simulated energy savings 31--36\% vs. simulated baseline \\

% done - checked
36 & \cite{Zhuang2018-fj} \newline 2018 & MPC & Sichuan Province, China & six above-ground floors of a shopping mall & forced-air with chillers, variable speed drivers, AHUs, and fan coils & 1 & minimize energy & 16\% energy savings vs. measured baseline \\

% done - checked
37 & \cite{Michailidis2018-zl} \newline 2018 & RL & Aachen, Germany & three conference rooms in a university research building & hybrid forced-air and radiant system with concrete core and air chillers & 5 & minimize energy & 34.7\% energy savings vs. measured baseline \\

% done - checked
38 & \cite{Lindelof2018-wt} \newline 2018 & MPC & Sargans, Switzerland & three zones within an office building & hydronic system with oil-fired boiler & 46 & minimize energy & 31.9\% energy savings vs. measured baseline \\

% done - checked
39 & \cite{Viot2018-sd} \newline 2018 & MPC & Bordeaux, France & conference room in a university building & radiant underfloor heating with AHU and fan coil unit for ventilation and backup heating & 30 & minimize energy & 30--40\% energy savings vs. simulated baseline \\

% done - checked
40 & \cite{Granderson2018-gp} \newline 2018 & MPC & Long Beach, CA; Dayton, OH; NYC; Washington, DC, US & office, courthouse, hospital, and high school buildings & forced-air variable frequency drive systems & 108--170 & minimize energy & -0.4--8.9\% energy savings vs. measured baseline \\

% done - checked
41 & \cite{Joe2018-lw} \newline 2018 & MPC & West Lafayette, IN, US & open-plan office space in a university research building & hydronic radiant floor cooling, air-cooled chiller assumed & 11 & minimize energy & 27\% energy savings vs. simulated baseline \\

% done - checked
42 & \cite{Kim2018-ku} \newline 2018 & MPC & West Lafayette, IN, US and FL, US & conference room in a university building and a retail store & forced-air systems with packaged RTUs & 3--60 & minimize energy and peak demand & 13--14\% energy savings and 14.5--50\% peak demand reduction vs. measured baseline \\

% done - checked
43 & \cite{Joe2019-ie} \newline 2019 & MPC & West Lafayette, IN, US & three open-plan office spaces in a university research building & radiant floor system in the first office, forced-air systems in the others & 10 & minimize energy and cost & 7.8--64\% energy savings and 34.7--78\% cost savings vs. measured and simulated baseline \\

% done - checked
44 & \cite{Lee2019-tj} \newline 2019 & MPC & West Lafayette, IN, US & open-plan office space in a university research building & hydronic radiant floor cooling, air-cooled chiller assumed & 19 & improve comfort and minimize energy & 23.1\% thermal dissatisfaction reduction with energy increase vs. measured baseline \\

% done - checked
45 & \cite{Zhang2019-ub} \newline 2019 & RL & Pittsburgh, PA, US & single-level office building & radiant heating system integrated with window mullions & 78 & minimize energy & 16.7\% energy savings vs. measured baseline \\

% done - checked
46 & \cite{Yang2019-xi} \newline 2019 & MPC & Singapore & experimental cells in a building performance test facility & hybrid system with active chilled beams and primary AHU-induced mixing & 1 & minimize energy & 14.7--20\% energy savings vs. measured baseline \\

% done - checked
47 & \cite{Yang2020-sj} \newline 2020 & MPC & Singapore & single lecture room in a university building & forced-air VAV with central plant and dedicated outdoor air system & 14 & minimize energy & 16--20\% energy savings vs. measured baseline \\

% done - checked
48 & \cite{Cotrufo2020-up} \newline 2020 & MPC & Varennes, QC, Canada & single-story building with workstations and meeting rooms & forced-air with electric boiler for heating and natural gas boilers as backup & 70 & minimize energy & 22.2\% energy savings vs. measured baseline \\

% done - checked
49 & \cite{Drgona2020-es} \newline 2020 & MPC & Hasselt, Belgium & three-floor office building, underground garage and under-roof apartment excluded & hybrid radiant and forced-air system with TABS and ground-source HP & 35 & minimize energy & 53.5\% energy savings vs. measured baseline \\

% done - checked
50 & \cite{Yang2020-js} \newline 2020 & MPC & Singapore & office and lecture theater in a multi-story institutional building & forced-air with ceiling-mounted VAV and external fan coil unit & 10 & minimize energy & 36.7--58.5\% energy savings vs. measured baseline \\

% done - checked
51 & \cite{Chen2020-rs} \newline 2020 & RL & Pittsburgh, PA, US & conference room in a university building & forced-air VAV with variable-speed airflow & 21 & minimize energy & 16.7\% energy savings vs. measured baseline \\

% done - checked
52 & \cite{Freund2021-nz} \newline 2021 & MPC & Hamburg, Germany & large government office building & radiant heating with district hot water & 90 & minimize energy & 30\% energy savings vs. measured baseline \\

% done - checked
53 & \cite{Sampaio2021-oe} \newline 2021 & MPC & Brussels, Belgium & five-floor building & forced-air with central heating, cooling equipment, and AHUs & 10--17 & minimize energy & 8.6--11.7\% energy savings vs. simulated baseline \\

% done - checked
54 & \cite{Yang2021-nc} \newline 2021 & MPC & Singapore & experimental cells in a building performance test facility & hybrid system with active chilled beams and primary AHU-induced mixing & 1--2 & minimize energy & 18.4--23.8\% energy savings vs. measured baseline \\

% done - checked
55 & \cite{Touzani2021-bx} \newline 2021 & RL & Berkeley, CA, US & office room in a controlled test facility & forced-air with variable-capacity AHU, shared chiller, PV, and battery & 7 & minimize cost & 0.6--39.6\% cost savings vs. measured baseline \\

% done - checked
56 & \cite{Jung2022-gy} \newline 2022 & RL & Seoul, South Korea & climate chamber in controlled environment & NR & $<1$ & improve comfort & 10.9\% predicted thermal discomfort (PPD) reduction vs. measured baseline \\

% done - checked
57 & \cite{Arroyo2022-yh} \newline 2022 & MPC & Heverlee, Belgium & unoccupied test room in a research facility & radiant heating via TABS, supplied by an electric boiler & 57--69 & compare MPC modeling approaches & no model consistently outperformed others vs. measured data \\

% done - checked
58 & \cite{KimBraun2022} \newline 2022 & MPC & West Lafayette, IN, US & conference room in a university building & forced-air with packaged RTUs & 5--15 & minimize cost & 3.6--8.7\% cost savings vs. measured baseline \\

% done - checked
59 & \cite{Blum2022-dc} \newline 2022 & MPC & Berkeley, CA, US & two office floors with closed offices below and open offices above & forced-air underfloor system with direct expansion cooling and heating & 31 & minimize energy & 40\% energy savings vs. measured baseline \\

% done - checked
60 & \cite{Naug2022-we} \newline 2022 & RL & Nashville, TN, US & three-story, energy-efficient mixed-use building & forced-air with a dedicated outdoor air system, VRF, and dual AHUs & 365 & minimize energy & 13.1--14.3\% energy savings vs. measured baseline \\

% done - checked
61 & \cite{Kimetal2022} \newline 2022 & MPC & Merced, CA, US & central cooling plant at a university & forced-air with chilled water, thermal storage, PV, and flexible chillers & 7 & maximize PV self-consumption, minimize peak demand and emissions & 26.4\% PV self-consumption increase, 9.8\% peak demand reduction, and 9.6\% emissions reduction vs. measured baseline \\

% done - checked
62 & \cite{Zhang2022-cm} \newline 2022 & MPC & Blue Lake Rancheria, CA, US & gas station convenience store & forced-air RTUs with refrigeration, freezer, battery, and PV integration & 365 & minimize cost and peak demand & 11.7\% cost savings and 34\% peak demand reduction vs. measured baseline \\

% done - checked
63 & \cite{Maddalena2022-nk} \newline 2022 & MPC & Campo Grande, Brazil & operating and waiting rooms in a hospital & forced-air cooling with AHUs and variable-speed chiller & 4 & minimize energy & experimental savings unreported; simulated energy savings 4.76\% vs. simulated baseline \\

% done - checked
64 & \cite{Lei2022-ef} \newline 2022 & RL & Singapore & office in a net-zero university building & hybrid system with dedicated outdoor air system and variable-speed ceiling fans & 10 & minimize energy & 14\% energy savings vs. measured baseline \\

% done - checked
65 & \cite{luo2022controlling} \newline 2022 & RL & US & university and mixed-use commercial buildings & hybrid system with chiller plant, AHUs, fan coils, and central air distribution & 90 & minimize energy & 9--13\% energy savings vs. measured baseline \\

% done - checked
66 & \cite{Zhan2023-hf} \newline 2023 & MPC & Singapore & six-zone office space in a net-zero energy building & forced-air with two fan coil units and individual zone VAVs & 49 & maximize PV self-consumption and building self-sufficiency & 19.5\% PV self-consumption increase and 10.6\% building self-sufficiency increase vs. simulated baseline \\

% done - checked
67 & \cite{Zhang2023-cf} \newline 2023 & MPC & West Lafayette, IN, US & two identical adjacent private office spaces & forced-air with individual VAV boxes and centralized AHU with reheat & 1 & minimize energy & 28--35\% energy savings vs. measured baseline \\

% done - checked
68 & \cite{Yue2023-ms} \newline 2023 & MPC & Zhejiang province, China & three-floor airport terminal & forced-air with air-cooled HPs, chilled water pumps, and AHUs & 15 & minimize energy & 10.6--37.3\% energy savings vs. measured baseline \\

% done - checked
69 & \cite{Ham2023-la} \newline 2023 & MPC & Costa Mesa, CA, US & six classrooms across two buildings in a K-12 school & forced-air with six RTUs serving each classroom & 30 & minimize peak demand & 22-–24\% peak demand reduction vs. measured baseline \\

% done - checked
70 & \cite{Lian2023-jx} \newline 2023 & MPC \newline (DeePC) & Lausanne, Switzerland & multipurpose academic building & forced-air HP & 4 & minimize energy & 18.4\% energy savings vs. measured baseline \\

% done - checked
71 & \cite{Saloux2023-cq} \newline 2023 & MPC & ON and QC, Canada & central heating plants & radiant steam system with non-condensing gas boilers for heating and hot water & 48--67 & minimize energy, cost, and GHG emissions & 1.3--2.8\% energy, cost, and GHG emissions reduction vs. measured baseline \\

% done - checked
72 & \cite{Stoffel2024-uu} \newline 2024 & MPC & Aachen, Germany & five university office spaces and a test hall in an energy facility & forced-air and radiant systems with district heating and concrete core activation & 14--33 & minimize energy & 5.3--22.2\% energy savings vs. measured baseline \\

% done - checked
73 & \cite{Ham2024-vg} \newline 2024 & MPC & NY, US & subspace in a small mixed-use commercial building & forced-air with mini-split HPs and a gas furnace & 13 & minimize cost and peak demand & 27\% cost savings and 23\% peak demand reduction vs. measured baseline \\

% done - checked
74 & \cite{Morovat2024-jr} \newline 2024 & MPC & Sainte-Marthe-sur-le-Lac, QC, Canada & six classrooms in a fully electric school building & hybrid with forced-air, radiant heating, HPs, and thermal storage & $<1$ & minimize cost and peak demand & 44\% cost savings and 47--95\% peak demand reduction vs. measured baseline \\

% done - checked
75 & \cite{Lolli2024-hx} \newline 2024 & MPC & Trondheim, Norway & a block of a 6-story office building & hydronic heating, mechanical ventilation, and sun-shading & 40 & minimize energy & 10\% energy savings vs. measured baseline \\

% done - checked
76 & \cite{Silvestri2024-kw} \newline 2024 & RL & Dübendorf, Switzerland & office in a modular high-performance research building & radiant hydronic TABS, hybrid VRF, and heat recovery ventilation & 49 & minimize energy and improve comfort & 68\% temperature violations reduction with similar energy consumption vs. simulated baseline \\

% done - checked
77 & \cite{Choi2024-yv} \newline 2024 & MPC & Incheon, South Korea & single-zone office space in a university building & forced-air direct expansion cooling system & 3 & minimize energy and cost & 25.2\% energy savings and 33.7\% cost savings vs. measured baseline \\

% done - checked
78 & \cite{Wang2024-kj} \newline 2024 & MPC, \newline RL & Syracuse, NY, US & office in a small-to-medium university building & forced-air AHU with VAV and variable-speed fan & 25--27 & minimize energy & 18--38.6\% energy savings vs. measured baseline \\

% done - checked
79 & \cite{He2025-fu} \newline 2025 & MPC & Xi'an, China & university research laboratory & forced-air with chiller, AHU, and VAV & 1 & minimize energy & 33.3--39.1\% energy savings vs. measured baseline \\

% done - checked
80 & \cite{Shi2025-jt} \newline 2024 & MPC \newline (DeePC) & Lausanne, Switzerland & multipurpose academic building & forced-air with rooftop unit and continuous ventilation & 12 & frequency regulation evaluation and minimize cost & improved frequency regulation and 24.6--28.7\% cost savings vs. measured baseline \\

\end{longtable}
\end{landscape}
%\twocolumn % Switch back to two-column mode

\end{document}